\documentclass{aa}

\usepackage{graphicx}
\usepackage{amsmath,epsfig,rotating,amssymb}

\begin{document}

\author{ Patrick Hennebelle\inst{1}}

\institute{ Laboratoire de radioastronomie, UMR 8112 du CNRS, 
\newline {\'E}cole normale sup{\'e}rieure et Observatoire de Paris, 24 rue Lhomond,
\newline 75231 Paris cedex 05,
France 
}



\titlerunning{ Star formation within clusters}
\authorrunning{Hennebelle}

\title{Formation of proto-clusters and star formation within clusters: 
apparent universality of the initial mass function ?}

\abstract{It is believed that the majority of stars form in clusters.
Therefore it is likely that the gas physical conditions that prevail
in forming clusters largely determine the properties of stars
that form  in particular, the initial mass function.} 
{We develop an analytical model to account for the formation
of low-mass clusters and the formation of stars within clusters. }
{ The 
formation of clusters is determined by an accretion rate, the 
virial equilibrium as well as  energy and thermal balance.
For this latter, both molecular and dust cooling are  considered
using published rates.    
The star distribution is computed within the cluster using the 
physical conditions inferred from this model and the Hennebelle \& Chabrier
theory.} {Our model reproduces well the mass-size relation of low mass
clusters (up to few $\simeq 10^3$ M$_\odot$ of stars corresponding 
to about five times more gas) and an initial 
mass function that is $i)$ very close to the Chabrier IMF,  
$ii)$  weakly dependent on the mass of the clusters, 
$iii)$ relatively robust to 
(i.e. not too steeply dependent on) variations of physical quantities such
as accretion rate, radiation, and cosmic ray abundances.} {The weak dependence
of the mass distribution of stars on the cluster mass results 
from the compensation between varying clusters 
densities, velocity dispersions, and 
temperatures that are all inferred from first physical principles. 
This constitutes a possible explanation for the 
apparent universality of the IMF within the Galaxy
although variations with the local conditions may certainly be observed.}


\maketitle

\section{Introduction}

Since the pioneering work of 
Salpeter (1955), the origin of the initial mass function 
(IMF; e.g. Kroupa 2002, Chabrier 2003) remains one 
of the most fundamental questions in astrophysics.
More particularly, its apparent universality, 
 Bastian et al. (2010), remains  debated 
and challenging.  Indeed, as described by 
these authors, the IMF has now been determined 
in many environments, and although 
some of the results may be interpreted as 
evidence of variations, neither systematic, nor 
undisputable evidence for such variations of the IMF are unambiguously 
reported. Therefore even if, as it will probably eventually turn out, 
some variations are finally clearly established,  in many environments, the 
variations of the IMF  remains  limited.  

Various theories have been proposed to explain the 
apparent constancy of the IMF. These theories often 
rely on the independence of the Jeans mass on the density.
For example, Elmegreen et al. (2008) computed the 
gas temperature in various environments and found a very weak
dependence of the Jeans mass  on the density since
the latter increases with temperature. 
Bate (2009) and Krumholz (2011) estimated 
the temperature in a massive collapsing clump in which the 
gas is heated by the radiative feedback due to accretion 
onto the protostars. Bate (2009) found that {\rm in the vicinity 
of the protostars, the Jeans mass
is typically proportional to $\rho^{1/5}$. While this later 
idea is interesting, it raises a few questions. First, 
the first generation of stars is, at least at the beginning
of the process, not influenced by the radiative feedback. 
Second, even when the stars start forming, the 
regions in which heating is important remain limited to the 
neighborhood of the protostars.  Thus it is not yet
demonstrated that most stars will be affected 
by this effect and that it is sufficient to make the IMF 
universal. Moreover the recent simulations by 
Krumholz et al. (2012) suggest that indeed the 
gas temperature is much too high when radiative 
feedback is included, leading to an IMF that is inconsistently 
shifted toward high masses. On the contrary when 
outflows are included, the radiation can escape along the 
outflow cavities
and they obtain IMF which are close to the observed one.

Another type of arguments invokes the variation 
of the effective polytropic index, $\gamma$, which 
in particular presents a local minimum at about $10^5$
cm$^{-3}$ (e.g. Larson 1985) due to the transition 
between cooling dominated lines and dust. This in particular 
has been proposed by Bonnell et al.(2006) and Jappsen et al. (2005). 
However, it remains unclear that it is actually the case because
the various simulations did not clearly establish
that changing the cloud initial conditions while 
keeping a fixed equation of state would lead to a CMF peaking 
at the same mass. Moreover Hennebelle \& Chabrier (2009) 
compare their predictions with the peak position found 
in numerical simulations of Jappsen et al. (2005) and find a good
agreement. Yet in the Hennebelle \& Chabrier theory, 
a local minimum of $\gamma$ does not determine  the peak of the 
CMF which still depends on the Mach number
for example.

More generally, these theoretical arguments assume that the Jeans 
mass is the only parameter that determines the IMF. 
This sounds rather unlikely because a distribution 
like the IMF is not entirely determined by a single parameter 
(peak position, width, and slope at high masses), 
moreover, analytical theories like the one proposed 
by Hennebelle \& Chabrier (2008, HC2008) and Hopkins (2012) 
explicitly depend on the Mach number. 
Although no systematic exploration of the Mach number
influence on the core mass function 
has been performed in numerical simulations, 
Schmidt et al. (2010) have explored the role of the forcing 
of the turbulence. They showed in particular that the core mass function
(CMF) is quite different when the forcing is applied in pure 
 compressible or pure  solenoidal modes. 
This clearly suggests that the CMF is affected by 
the velocity field. Even more quantitatively, Schmidt et 
al. (2010) found a good agreement between the CMF they measure
and the analytical model of HC2008
 which seemingly confirms  the Mach number 
dependence of this model.
From a physical point of view, it is well established that 
the density probability distribution function
(PDF) is strongly related to the Mach number 
(V\'azquez-Semadeni 1994, Padoan et al. 1997, 
Passot \& V\'azquez-Semadeni 1998, Kim \& Ryu 2005, 
Kritsuk et al. 2007, Federrath et al. 2008, 
Audit \& Hennebelle 2010). 
Accordingly, because
 the density PDF is clearly important with 
regard to the 
Jeans mass distribution within the cloud, it would be quite 
surprising if the Mach number had no influence on the 
CMF. 

Another line of explanation regarding the universality of the IMF 
has been proposed by HC2008, who argue that 
because the peak position of the IMF is proportional to  the mean Jeans 
mass and inversely proportional to  the Mach number, there is 
a compensation because from Larson relations, 
the density decreases when the velocity dispersion increases
and thus the mean Jeans mass and the Mach number increase
 at the same time (see Eq.~47 of HC2008 and figure~8 of HC2009). 
One problem  of this explanation is, however, 
that Larson 
relations present a high dispersion and that there are clouds for example, with the same density but different Mach numbers
that may therefore lead to different IMF in particular at low 
masses. 

Generally speaking, all approaches that attempt 
to understand the universality of the IMF suffer from the 
variability of the star-forming cloud conditions. 
This clearly emphasizes the need for a better
understanding of the physical  conditions under which 
stars form. In this respect, an important key is 
 that most stars (say 50-70\%) seem to 
form in clusters (Lada \& Lada 2003, Allen et al. 2007), 
which is a strong
motivation to study  the formation of clusters. 
Note, however, that Bressert et al. (2010) moderate
this picture to some extent.

  A word of caution is nonetheless necessary here. 
The constancy of the IMF is debated and some observations
even in the Galaxy may indicate that some variability has
already been observed (see e.g. Cappellari et al. 2011 for 
early types galaxies). For a discussion and 
good summary on this problem we refer the reader to Dib et al. (2010).
Moreover, these authors have developed an analytical 
model of the core formation within  proto-clusters that 
takes into account the  turbulent formation of 
dense cores as well as their accretion of gas that could
modify the CMF and lead to IMF variability. Along a 
similar line, Dib et al. (2007) showed how the coalescence
of cores can explain the development of a 
top-heavy IMF.

In this paper, we first develop an analytical model 
for the formation of clusters, more precisely, the formation 
of proto-clusters, i.e. the gas dominated phase that 
eventually leads to  star-dominated clusters.
Our model relies on the gas accretion onto the 
proto-clusters from  parent clumps that feed them in mass and in energy.
This allows us to predict the physical quantities such as
radius, mean density, and velocity dispersion
within the proto-clusters as a function of the accretion rate. 
By comparing with the data of embedded clusters from 
Lada \& Lada (2003), we can estimate the accretion rate onto 
these proto-clusters and verify that our model fits the 
observational data well. In a second step, we calculate 
the gas temperature within the proto-clusters by computing 
the various heating and cooling, and we apply 
a time-dependent version of 
the model of HC2008 to predict the mass spectrum of the 
self-gravitating condensations and in particular 
to study their variability with 
proto-cluster masses and accretion rates.

The second part of the paper presents the  analytical model
of the proto-clusters and the comparison with the 
observational data. The third part is devoted to the 
calculation of  
the thermal balance as well as to the description of the 
HC2008 theory. In the fourth part, we calculate 
the mass spectra within the  proto-clusters. The fifth section 
concludes the paper.

\section{Analytical model for low mass cluster formation}
Clusters are likely to play an important role for the 
formation of stars in the Milky Way (e.g. Lada \& Lada 2003) 
and probably for most galaxies. 
It seems therefore a necessity, in order to understand how star forms to 
obtain a good description of the gas physical conditions within proto-clusters. 
For that purpose, we develop here an analytical model that is based 
on the following general ideas.
First, proto-clusters are initially likely to be gravitationally bound entities.
However, proto-clusters are not  collapsing, which means that a support is actually 
compensating gravity, which we assume is the turbulent dispersion.
 This will be expressed by applying the virial theorem to the proto-cluster. Because turbulence
is continuously decaying with a characteristic time on the order of
the crossing time, it must be continuously sustained. We assume that 
the continuous accretion of gas into the proto-cluster from the parent clump is 
the source of energy. Indeed, accretion-driven turbulence has been recently proposed 
to be at play  in various contexts (Klessen \& Hennebelle 2010, Goldbaum et al. 2011). 
Our first step is therefore to discuss the parent clumps and the resulting accretion rate
onto the proto-cluster. Once an accretion rate is inferred, an energy balance can be written
and together with the relation obtained from the virial theorem lead
to a link between the mass, the radius, and the velocity dispersion within the cluster.

\subsection{Parent cloud, protocluster and accretion rate}

\subsubsection{Definitions and assumptions}

Let us consider a clump of mass $M_c \simeq 10^3-10^6 M_\odot$ and radius $R_c$ that 
follows Larson relations (Larson 1981, Falgarone et al. 2004, Falgarone et al. 2009):
\begin{eqnarray}
n_c = n_0  \left( { R_c \over 1 {\rm pc}} \right)^{-\eta_d},\,\,
\sigma_{\rm rms} = \sigma_0  \left( { R_c \over 1 {\rm pc}} \right)^{\eta},
\label{larson}
\end{eqnarray}
where $n_c$ is the clump gas density and $\sigma_{rms}$ the internal rms velocity. 
The exact values of the various coefficients remain somewhat uncertain. 
Originally, Larson (1981) estimate $\eta_d \simeq 1.1$ and $\eta \simeq 0.38$,
but more recent estimates (Falgarone et al. 2004, 2009) using 
larger sets of data suggest that $\eta_d \simeq 0.7$ and $\eta \simeq 0.5$.
These later values  agree well with the estimate from numerical 
simulations of supersonic turbulence (e.g. Kritsuk et al. 2007) and we
therefore use them throughout the paper although we will compare 
the accretion rates obtained for both sets of values.
 
As explained above, we consider the formation of a cluster 
within the parent clump. Because we are essentially 
considering the early phase during which the gas still dominates
over the stars, we refer to it as the proto-cluster. 
Let $M_*$ be its mass and  $\alpha_{*,c}=M_c/M_*$ be
 the ratio between the parent clump and proto-cluster masses.
Let us stress that $M_*$ is the mass of gas within the proto-clusters and not 
 the mass of stars. Throughout this work the star component is not considered.

Obviously, the proto-cluster is accreting from the parent clump, which 
is therefore regulating the rate at which matter is delivered 
onto the proto-cluster. The immediate 
question concerns then  
the value of the accretion rate ? To answer this
we consider two different approaches. First we 
assume that the proto-cluster clump is undergoing Bondi-type accretion, 
i.e. accretion regulated by its own gravity. However because
this process can be sustained only if sufficient matter
 is  available for accretion, we also explore the possibility 
that the accretion onto the proto-cluster is regulated by the 
accretion onto the parent clumps. In this case, the 
accretion is  due to the mechanism that is at the origin of the 
clump formation and that also sets the Larson relations. 
Both assumptions lead to similar, although not identical values and  
dependence.

\subsubsection{Bondi-type accretion rate}
The Bondi-Hoyle accretion rate that an object of mass $M_*$
is experiencing in a medium of density $\rho$ and sound speed $c_s$
is given by $\dot{M}=4 \pi \rho G^2 M_*^2 / c_s^3$ (Bondi 1941).
In a turbulent medium the accretion rate is not well established.
While a simple expression is  given by $\dot{M}=4 \pi \rho G^2 M_*^2 
/ (c_s^2 + \sigma^2)^{3/2}$, where 
$\sigma$ is the velocity dispersion, Krumholz et al. (2005) obtain
and discuss more refined estimates.
We stress that these estimates assume that the accreting reservoir is 
infinite, which, as already mentioned may not be true in the present
context. Moreover, these accretion rates are valid only for point 
masses and it is unclear whether they apply to more 
extended objects such as proto-clusters. 

To estimate an accretion rate, here we  used
the modified Bondi expression.  
The density $\rho$ and the velocity dispersion $\sigma$ are 
 those from the parent clump stated in Eqs.~(\ref{larson}). 
Because these quantities depend on the parent clump mass, $M_c$, 
and because $M_*$ likely increases with $M_c$, it 
is clear that the dependence of the accretion rate is not 
proportional to $M_*^2$ but rather has a more shallow dependence
essentially because $\rho$ and $\sigma$  decreases
and increases respectively when $M_c$ increases. To arrive at a more quantitative 
estimate, we assume that $\alpha_{*,c}$ 
does not depend on $M_c$ and is therefore constant. We obtain
\begin{eqnarray}
\nonumber
\dot{M} &=&  { 4 \pi \rho G^2 M_*^2 \over 
 \sigma^3 } = { 4 \pi M_p n_0  G^2 M_c^2 \over 
 \alpha_{*,c}^2 \sigma_0^3 (R / 1 {\rm pc})^{\eta_d+3 \eta}} \\
& \propto& 
 \alpha_{*,c}^{-2} M_c^{ 3(2 -(\eta_d+ \eta)) \over (3-\eta_d) }.
\label{accret_bond} 
\end{eqnarray}
For $\eta_d=0.7$, $\eta=0.45$, we get $\dot{M} \propto M^{-1.1}$
while for $\eta_d=1.1$, $\eta=0.4$, we obtain $\dot{M} \propto M^{-0.8}$.
Figure~\ref{accret_rate} shows the resulting accretion rate, 
BL1 and BL2 referring  to the first and second sets of parameters respectively,
the values $\alpha_{*,c}=3$ (upper curves) and  
$\alpha_{*,c}=5$ (lower curves)
were used. This likely 
corresponds to upper values of the accretion rate because higher 
values of $\alpha_{*,c}$  lead to lower values of 
$\dot{M}$, while for  lower values of $\alpha_{*,c}$,
 the mass within the parent clumps
is equal to or even smaller than, the mass within the protocluster and it is
therefore extremely unlikely that Bondi-type accretion is relevant at all.

\begin{figure} 
\includegraphics[width=9cm]{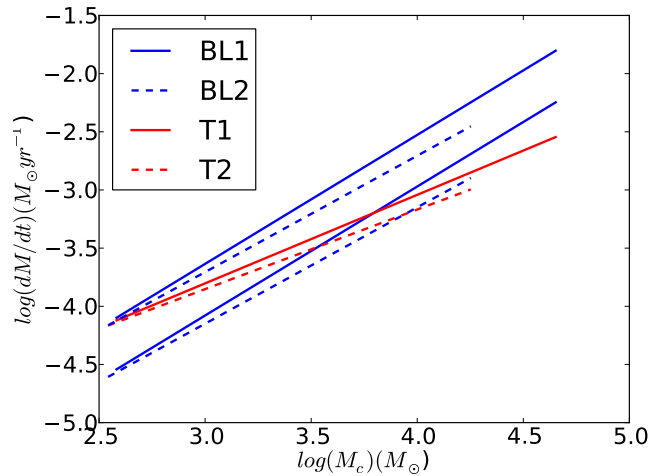}
\caption{Accretion rate onto the proto-cluster  against clump mass.
BL1 and BL2 represent Bondi-Lyttleton accretion for 
$\eta_d=1.1$, $\eta=0.4$ and $\eta_d=0.7$, $\eta=0.45$, respectively.
T1 and T2 represent the turbulent accretion rate with the same values. 
Upper (lower) BL curves are for parent clump over proto-cluster
mass of 3 (5).}
\label{accret_rate}
\end{figure}

\subsubsection{Turbulent accretion rate}
The second type of accretion we consider is the one that 
is responsible for the clump formation. 
The exact nature of these clumps is still a matter of debate 
but simulations have been reasonably successful to explain at least the low-mass part of the distribution (e.g. Hennebelle \& Audit 2007, 
Banerjee et al. 2009, 
Klessen \& Hennebelle 2010) in which they can essentially be seen
as turbulent fluctuations. In this view, as the properties of the 
clumps stated by Eqs.~(\ref{larson}) 
are the result of the turbulent cascade, that is to say, the 
material within the clumps is delivered by the compressive motions of 
the surrounding diffuse gas, it seems reasonable to construct an accretion 
rate out of the Larson relations.

The clump crossing time, $\tau_c$ is about 
$2 R / (\sigma / \sqrt{3})$ where 
 the factor $\sqrt{3}$ accounts for 
the fact that it is the one-dimensional velocity dispersion 
that is relevant 
for estimating the crossing time. 

The accretion rate of diffuse gas onto the clump is expected to be
\begin{eqnarray}
\nonumber
\dot{M} &\simeq& {M_c \over \tau_c} \simeq {M_c \over 2 R_c/(\sigma_{rms}/\sqrt{3})},  \\ \label{accretion_rate}
& \simeq & M_c^{{2-\eta_d + \eta \over 3 - \eta_d}} 
{ \sigma_0 \over 2 \sqrt{3}} 
 \left( {4 \pi \over 3} n_0 m_p\right) ^{ 1 - \eta \over 3 - \eta_d}
(1 \, {\rm pc})^{ \eta_d - 3 \eta \over 3 - \eta_d} ,
\label{dM_dt} \\
&=&  \dot{m}_4 \left( {M_c \over 10^4 M_\odot} \right) ^{\eta_{acc}},
\nonumber \\
&\simeq&  9 \times 10^{-4} M_\odot \, {\rm yr}^{-1} \,  \left( {M_c \over 10^4 M_\odot} \right) ^{2 - \eta_d + \eta \over 3-\eta_d}  \nonumber \\
&& \times {\sigma_0 \over 0.8 \, {\rm km \, s^{-1}} } \left( { n_0 \over 1000 \, {\rm cm}^{-3}} \right)^{1 - \eta \over 3 - \eta_d},
\nonumber 
\end{eqnarray}
where $m_p$ is the mass per particle. Note that to estimate 
$\dot{m}_4$ we  used $\eta_d=0.7$ and $\eta=0.45$.

The Larson relations that describe the mean properties of the clump
in which the cluster forms are likely a direct consequence of
 the interstellar turbulence.
Thus, the accretion is likely to last a few clump crossing times, which 
is typically the correlation  time in turbulence.

If the clump is sufficiently dense, it  undergoes gravitational collapse and 
forms a cluster of mass $M_*$. Here we attempt to understand the cluster formation 
phase. For that purpose, we assume that:
\begin{description}
\item[-] as already explained it starts with a gas-dominated phase, i.e. the mass of the cluster is  dominated
 by the mass of its gas and not its stars
\item[-] this phase is  quasi-stationary, i.e. the cluster properties evolve slowly with respect to its crossing time.
\item[-] the accretion rate onto the cluster is comparable
 to the accretion rate of diffuse gas onto the clump 
and is therefore close to the value given by Eq.~(\ref{dM_dt}).
\end{description}
 Of these three assumptions,  the second one is certainly the 
least obvious. Indeed, initially no star will have formed by definition,
while what we assume in practice regarding the accretion rate is 
simply that it scales as $\dot{M} \propto M^{\eta_{acc}}$ and 
we fit the coefficient using observed proto-clusters.
Taking time-dependence into account would certainly be highly desirable 
at some stage but will probably
 not modify the results very significantly.

It is useful to express the accretion rate as a function of the proto-cluster mass and 
we write
\begin{eqnarray}
\dot{M} =  \dot{m}_4 \left( { \alpha_{*,c} \over  10^4 M_\odot} \right) ^{\eta_{acc}}  M_* ^{\eta_{acc}} = Q_0 M_* ^{\eta_{acc}}.
\label{accretion}
\end{eqnarray}
Assuming that the parameter $\dot{m}_4$ stays constant, this 
implies that the accretion rate onto the proto-clusters is controlled
by  $M_c$, the mass of their parent clumps, and thus by  $\alpha_{*,c}=M_c/M_*$,
which  effectively quantifies the 
strength of the accretion onto the proto-cluster. In practice, 
$\dot{m}_4$ could vary while $\alpha_{*,c}$ would be 
constant, it is  equivalent, however, and does not make any difference if one uses one 
or the other. 
Below we use $\alpha_{*,c}$ to  quantify the accretion onto the 
proto-cluster.

\subsection{Virial equilibrium}
Because a cluster  initially is a bound system, which unlike a protostellar core
is not globally collapsing, it seems clear that on large scale some sort 
of mechanical equilibrium is established, to
describe which we use the virial theorem.  
When applying the virial theorem to the cluster, it must be taken into account
 that it is gaining mass by accretion and is therefore not a close system. 
In Appendix A, we show that when applied to an accreting system, the 
expression of the virial theorem becomes
\begin{eqnarray}
 \dot{M}^2 {d R_*^2 \over dM_*}   + M_* \sigma_*^2 - 3 P_{\rm ram} V_* + E_g =0.
\label{virial}
\end{eqnarray}
In this expression, $M_*$ is the cluster mass, $\sigma_*$ is the rms velocity 
of the gas within the cluster, $V_*=4 \pi / 3 R_*^3$ is  its volume, 
and $P_{\rm ram}$ is the external pressure exerted by the infalling clump gas onto the 
cluster which 
dominates over the thermal pressure. Assuming that  the  gas within the parent clumps is gravitionally attracted by the proto-cluster, the 
 infall velocity is simply 
the gravitational freefall, and we obtain
\begin{eqnarray}
 v _{\rm inf}  = \sqrt{2 G M_* \over R_*}. 
\label{infall}
\end{eqnarray}
The ram pressure is given by $P_{\rm ram} = \rho_{\rm inf} v_{\rm inf}^2$ while the 
accretion rate $\dot{M}$ leads to the relation $\dot{M} = 4 \pi R_*^2 \rho_{\rm inf} v_{\rm inf}$,
therefore leading to 
\begin{eqnarray}
P_{\rm ram} = {\dot{M} \over 4 \pi R_*^2} v_{\rm inf}  = {G^{1/2} \dot{M} M_* ^{1/2} \over 2 \sqrt{2} \pi R_*^{5/2}}.
\label{ram}
\end{eqnarray}

The gravitational energy of a uniform density sphere is well-known to be 
$-(3/5) G M_*^2 / R_*$, and thus we obtain
\begin{eqnarray}
 \dot{M}^2 {d R_*^2 \over dM_*}   + M_* \sigma_*^2 - 3 P_{\rm ram} V_* - {3 \over 5}
{G M_*^2 \over R_*}  =0.
\label{virial}
\end{eqnarray}

\subsection{Energy balance}
\label{balance}
As stated in Eq.~(\ref{virial}), the proto-cluster is confined by its own gravity and by the ram 
pressure of the incoming flow.  The velocity dispersion of the gas it contains resists these 
two confining agents. To estimate its magnitude, we assume an
  equilibrium between energy injection and turbulent dissipation
\begin{eqnarray}
{M_* \sigma_*^2 \over 2 \tau _{\rm cct}} \simeq \dot{E}_{\rm ext} + \dot{E}_{\rm int},
\label{energy}
\end{eqnarray}
where $\tau _{\rm cct}$ is the cluster crossing time and is about $\tau _{\rm cct} \simeq 2 R_* / {\sigma_* / \sqrt{3}}  $.
$\dot{E}_{\rm int}$ and $\dot{E}_{\rm ext}$ are  the external and 
internal source of 
energy injection which compensate the energy dissipation stated by the left side. Note that in principle a complete energy equation could be 
written that would entail thermal energy but also terms taking 
into account the expansion or contraction of the proto-cluster (see e.g. 
Goldbaum et al. 2011). However, the exact amount of energy 
dissipated by turbulence is not known to better than a factor of a few and 
the same is true regarding the efficiency with which the accreting gas 
can sustain turbulence (giving that a certain fraction can 
 quickly dissipate in shocks). Since these two contributions 
are dominant, we consider a simple balance at this stage
from which physical insight can be gained.

From Eq.~(\ref{energy}), we obtain
\begin{eqnarray}
 \sigma_*^2  \simeq  \left( 2 \sqrt{3} R_* {\dot{E}_{\rm ext} + \dot{E}_{\rm int} \over M_*} \right)^{2/3}.
\label{energy2}
\end{eqnarray}
Note that in this work we do not consider any internal energy source such as 
supernova explosions, jets, winds, and ionizing radiation from massive stars, accordingly
$\dot{E}_{\rm int}=0$ is assumed in this paper. 

The external source of energy is generated by
 the accretion process itself because
 the kinetic energy 
of the infalling material triggers motion within the cluster. This energy flux is 
on the order of $(1 / 2) \dot{M} v_{\rm inf}^2 \simeq G \dot{M} M_* / R_* $.
It is  slightly larger than this value, however, because  this expression corresponds
to the kinetic energy of the gas when it reaches the cluster boundary.
In practice the gas accreted onto the cluster continues to fall inside the cluster 
 and gain additional energy. In appendix B, we infer the corresponding
value, whose expression is
\begin{eqnarray}
\dot{E}_{\rm ext} = {6 \over 5} {G \dot{M} M_* \over R_*} \simeq {G \dot{M} M_* \over R_*}. 
\label{energy_ext}
\end{eqnarray}

Thus with Eq.~(\ref{virial}),~(\ref{energy2}) and~(\ref{energy_ext})
we get 
\begin{eqnarray}
\nonumber
{3 \over 5}{G M_* \over R_*} + \sqrt{2} G^{1/2} \dot{M}   
{R_*^{1/2} \over M_*^{1/2}}  &=&  \left( K G \dot{M}  \right)^{2/3} \\ 
&& + {1 \over 2} {\dot{M}^2 \over M_*} {d R_* ^2 \over dM_*},
\label{master}
\end{eqnarray}
where $K=12 \sqrt{3}/5$ although as discussed above it 
suffers from large uncertainties.

\setlength{\unitlength}{1cm}
\begin{figure} 
\begin{picture} (0,13)
\put(0,7){\includegraphics[width=9cm]{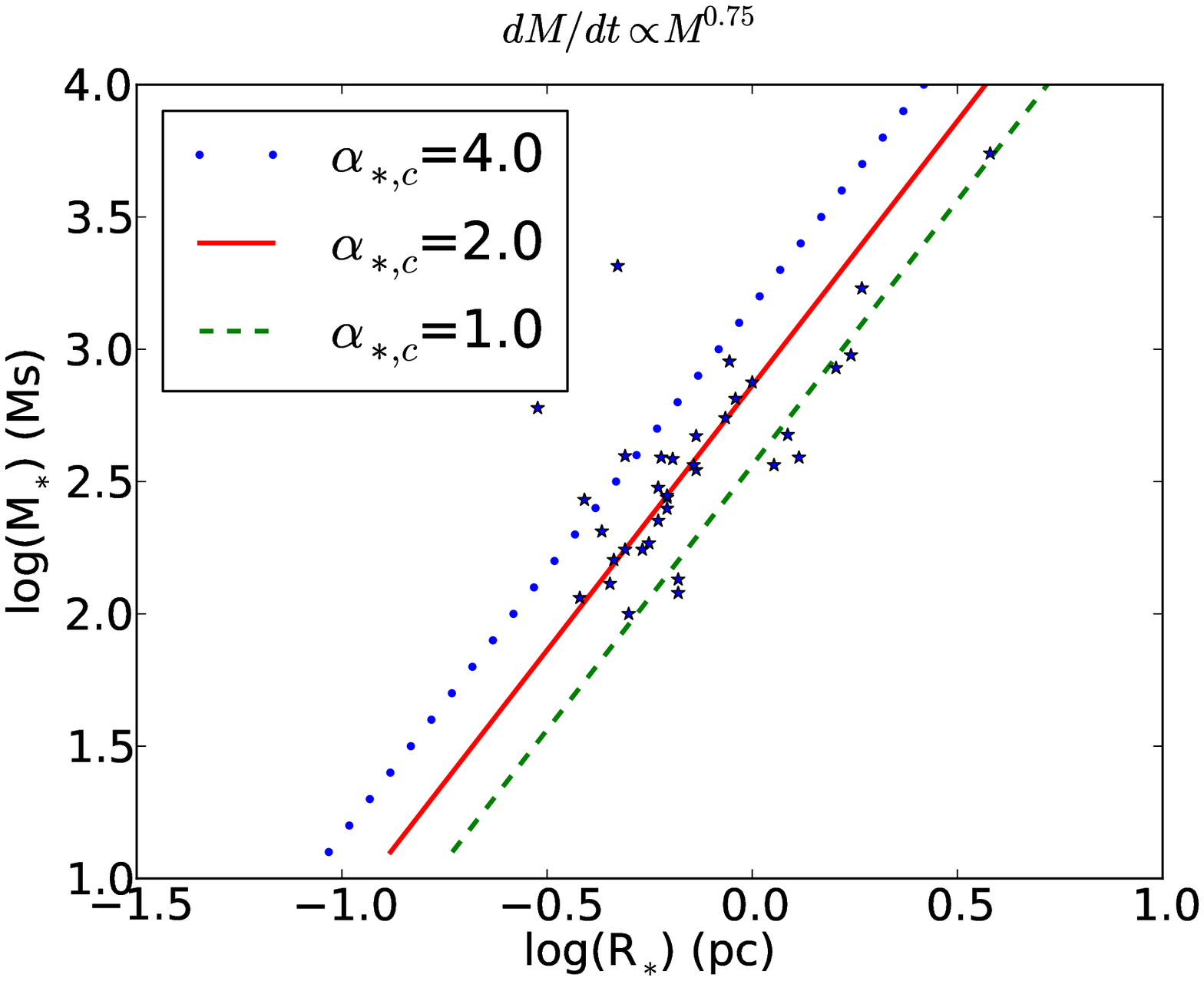}}
\put(0,0){\includegraphics[width=9cm]{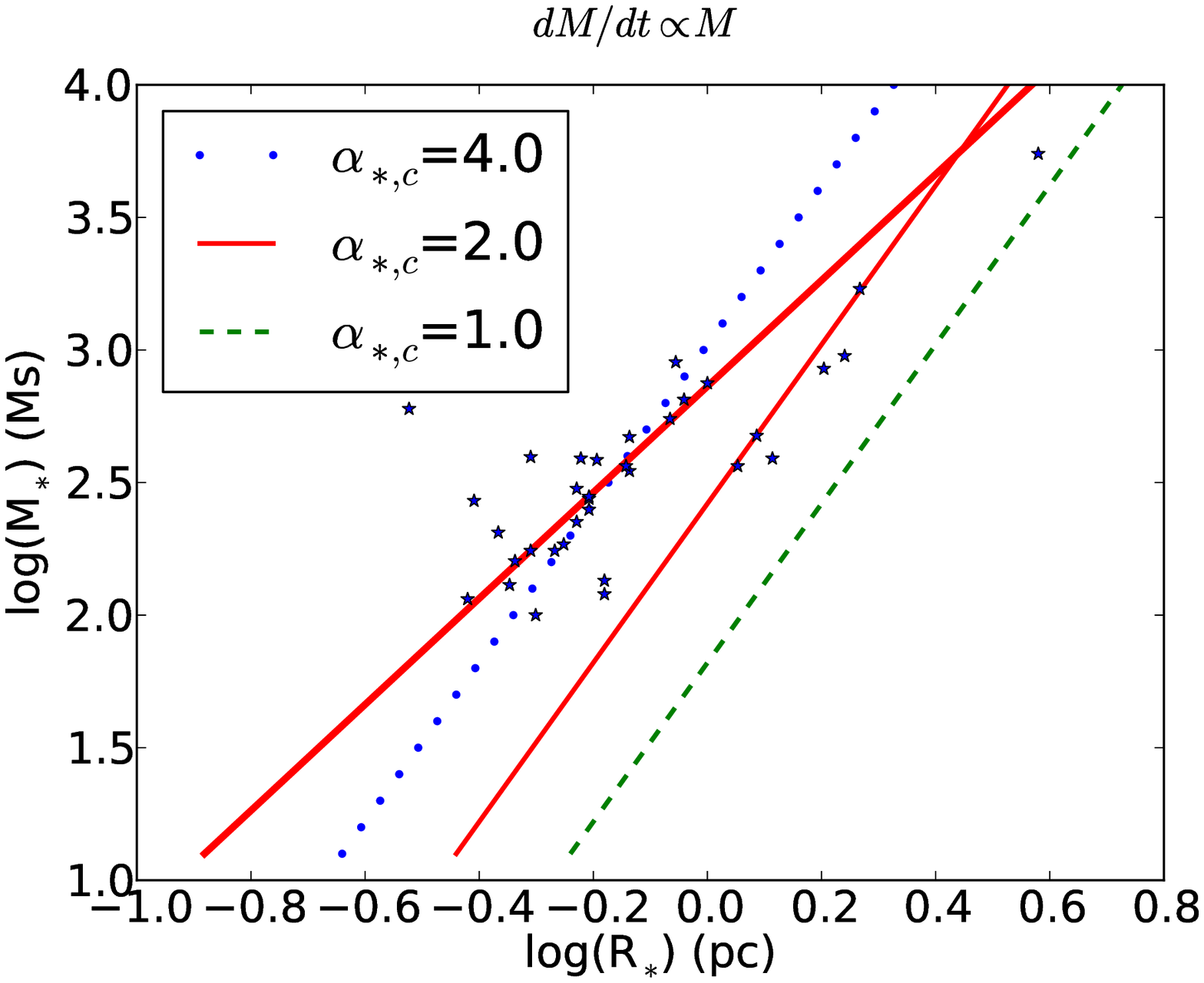}}
\end{picture}
\caption{Upper panel:
Mass of the cluster as a function of its radius as predicted by 
Eq.~(\ref{master}), i.e. for a turbulent type accretion for
which $\dot{M} \propto M^{0.75}$. 
Three values of $\alpha_{*,c}$ are displayed. 
The stars correspond to the embedded clusters listed in table 1 
of Lada \& Lada (2003) for which the mass has been multiplied by 5 
to account for a star formation efficiency of about 20$\%$.
Lower panel: same as lower panel for $\dot{M} \propto M$. 
The thick solid line corresponds to the solid line of the 
upper panel for comparison. }
\label{comp_lada}
\end{figure}

\subsection{Result: Mass vs radius of clusters}
\subsubsection{Simplified expression}
Before  deriving exact solutions of Eq.~(\ref{master}) and 
to obtain some physical hint, we start by discussing the simplified case
where the terms proportional to $P_{\rm ext}$ and $\dot{M}^2$ are ignored.
In this case we simply have
\begin{eqnarray}
{G M_* \over R_*}  \simeq  4.3 \left( G \dot{M} \right)^{2/3}.
\label{accret_M_R}
\end{eqnarray}
Using Eq.~(\ref{dM_dt}), we obtain
\begin{eqnarray}
\nonumber
 M_*   &\simeq&  4.3^{1 \over 1-2 \eta_{acc}/3 } 
 \left( G^{-1}  (\dot{m}_4) ^2  
\left( {\alpha_{*,c} \over 10^4 M_\odot} \right)^{2 \eta_{acc}} \right)^
{1 \over 3-2 \eta_{acc}}  \\
&  \times&
(R_*) ^{1/(1-2 \eta_{acc}/3)},
\label{accret_M_R}
\end{eqnarray}
where we recall that $ \alpha_{*,c} = M_c / M_*$ is the ratio of cluster over clump masses.
As seen from Eq.~(\ref{dM_dt}), $\eta_{acc} \simeq 0.75$ ($\eta=0.4-0.5$), therefore ${1/(1-2 \eta_{acc}/3)} \simeq 2$, which is 
the value that we adopt to perform this simplified calculation. We obtain
\begin{eqnarray}
 M_*   \simeq   4.3^{2} G^{-2/3}  (\dot{m}_4) ^{4/3}  (\alpha_{*,c} / 10^4 M_\odot)   R_* ^{2},
\label{accret_M_R}
\end{eqnarray}
which implies
\begin{eqnarray}
 M_*    \simeq  730 \; M_\odot  \times \alpha_{*,c}    \left( {\dot{m}_4 \over 10^{-3} M_\odot {\rm yr}^{-1} }\right) ^{4/3}  
  \left(  {R_* \over 1 {\rm pc}} \right) ^{2}.
\label{accret_M_R}
\end{eqnarray}

As suggested by Eq.~(\ref{accretion_rate}), the value of $\dot{m}_4$ is 
expected to be about a few 
$10^{-3} M_\odot {\rm yr}^{-1}$  while assuming that the 
mass of proto-cluster is comparable but smaller than the mass 
of its parent clump,
$\alpha _{*,c}$ which is likely to be about 1-4. 
Accordingly, for a 1 pc cluster, one typically
expects  a mass of about 700-2000 $M_\odot$ of gas. 

\subsubsection{Complete expression  for turbulent-type accretion}
We now solve Eq.~(\ref{master}) in the general case, i.e. without 
neglecting some of the terms.
It can be shown that assuming $\eta_{acc}=0.75$, 
the solution of this equation can still be written 
as $M_*  = M_*^0 (R_* / 1 {\rm pc})^2$., where $M_*^0$ satisfies
\begin{eqnarray}
\nonumber
{3 \over 5} G M_*^0 &+& \sqrt{2} G^{1/2} Q_0'  (M_*^0) ^{1/4}  = \\
\nonumber
& & \left( 12 \sqrt{3}/ 5 G Q_0'  \right)^{2/3} \sqrt{M_*^0}
+ {1 \over 2} (Q_0')  ^2 {1 \over \sqrt{M_*^0}}, \\
Q_0' &=& \dot{m}_4  \left( {\alpha_{*,c} \over 10^4 M_\odot} \right)^{0.75}
\times (1 {\rm pc})^{3/2}.
\label{eq_alpha}
\end{eqnarray}
Solving this equation numerically for $\alpha_{*,c}$=1, 2 and 4, 
we infer 
$M_*^0 \simeq 4144, \, 1381 $ and 414 M$_\odot$, respectively,  which is 
about 1.8 times lower than the value estimated in Eq.~(\ref{accret_M_R}).

We note that with $M_* \propto R_* ^2$, we derive that
the gas density and the column density  follow  
$n_* \propto R_*^{-1}$, respectively, while $\Sigma$ is constant. Quantitatively,
we have the three relations

\begin{eqnarray}
\nonumber
M_*   &=& M_*^0 
 \left( {R_* \over 1 {\rm pc}} \right)^2, \\
\label{cluster_rel}
n_*  &\simeq& 4600 \, {\rm cm}^{-3} 
{M_*^0 \over 10^3 M_\odot} \left( {R_* \over 1 {\rm pc}} \right)^{-1}, \\
\Sigma_{*,c}  &=& 2.8 \times 10^{22} {\rm cm}^{-2} {M_*^0 \over 10^3 M_\odot}, 
\nonumber
\end{eqnarray}
where $M_*^0$ is defined by Eq.~(\ref{eq_alpha}), 
$n_*= M_* / (4 \pi / 3 m_p R_*^3) $ and 
$\Sigma_{*,c} = 2 n_* R_*$.

Finally, from Eqs.~(\ref{accretion_rate}), (\ref{energy2}) and 
(\ref{energy_ext}), we obtain the internal gas velocity dispersion within
the proto-clusters
\begin{eqnarray}
\label{sigma}
\nonumber
\sigma _* &=& (2 \sqrt{3} G Q_0 (M_*^0)^{0.75})^{1/3} 
\left( {R_* \over 1 {\rm pc}} \right) ^{1/2}, \\
\nonumber
&=& (2 \sqrt{3} G)^{1/3} \dot{m}_4^{1/3} 
\left( {\alpha_{*,c} M_*^0 \over  10^4 M_\odot} \right)^{1/4} 
\left( {R_* \over 1 {\rm pc}} \right) ^{1/2}, \\
 &=& 1.4 \; {\rm km \, s}^{-1}
  \alpha_{*,c} ^{1/4} 
\left( {M_*^0 \over 10^3 M_\odot} \right)^{1/4} 
\left( {R_* \over 1 {\rm pc}} \right) ^{1/2}.
\label{cluster_sig}
\end{eqnarray}

\subsubsection{Bondi-type accretion}
\label{bondi-type}
As discussed above, combining Bondi-type accretion and Larson's 
relations, we infer accretion rates whose mass dependence
 follows $dM/dt \propto M^{\eta_{acc}}$ with $\eta_{acc} \simeq 1$.
Using this canonical value, it is easy to see 
that the solution of  Eq.~(\ref{master}) is $M_* = M_*^0 (R / (1 {\rm pc}))^3$,
where $M_*^0$ satisfies the following equation:

\begin{eqnarray}
\nonumber
{3 \over 5} G M_*^0 &+& \sqrt{2} G^{1/2} Q_0'  (M_*^0) ^{1/2}  = \\
\nonumber
& & \left( 12 \sqrt{3}/ 5 G Q_0'  \right)^{2/3} (M_*^0)^{2/3}
+ {1 \over 3} (Q_0')  ^2, \\
Q_0' &=& \dot{m}_4  \left( {\alpha_{*,c} \over 10^4 M_\odot} \right)^{0.75}
\times (1 {\rm pc})^{3/2}.
\label{eq_alpha}
\end{eqnarray}
That is to say, a dependence of the accretion rate
 ${\dot M} \propto M$ leads for the protoclusters to 
a mass-size relation $M \propto R^3$, i.e.  density 
is independent of the mass.

\subsubsection{Comparison with observations of embedded clusters}
For  canonical values of $\alpha_{*,c}=2$ and $\dot{m}_4=10^{-3}$ 
M$_\odot$~s$^{-1}$,  the relation 
$M_* \simeq 1380 \; {\rm M}_\odot \; (R_* / 1 {\rm pc})^2$ holds. To test 
our model, we compare it with the data of embedded clusters 
reported in table 1 of Lada \& Lada (2003). Because the mass quoted 
in this table is the mass of the stars, one must apply a correction 
factor to obtain the mass of the gas. In table 2, Lada \& Lada
give an estimate of this ratio for seven clusters. The star mass over gas mass
ratio is typically between 0.1 and 0.3 with an average value that is 
about 0.2.  Thus to perform our comparison, we simply multiply the 
star masses of table 1 by a factor of 5. By considering a unique 
star formation efficiency, we certainly increase the dispersion in the data. 
However since  a few values for the star formation efficiency are available, 
this is unavoidable.

The upper panel of Fig.~\ref{comp_lada} shows the results for three values of 
$\alpha_{*,c}$, the clump  over proto-cluster mass  ratio,
 namely 1, 2 and 4. 
Evidently, a good 
agreement is obtained with  $\alpha_{*,c} \simeq 2$
over almost 2 orders of magnitude in mass, 
although the dispersion of the observed values
is not negligible (factor 2). This may suggest that the relation stated by 
Eq.~(\ref{accretion_rate}) is relatively uniform throughout the 
Milky Way. Because this is likely a consequence of interstellar turbulence, 
it may  reflect the universality of its properties.
Note that as emphasized in Murray (2009), high-mass clusters 
($>10^4$ M$_\odot$) present a different mass-size relation. This may 
indicate that for more massive clusters other energy sources than stellar 
feedback should be considered.  It is also likely
that the feedback  strongly affects the proto-clusters 
when enough stars have formed. In particular, gas expulsion 
is likely to occur and feedback probably sets at least 
in part, the star formation efficiency (e.g. Dib et al. 2011). This late 
phase of evolution is  not addressed here, however. 

The lower panel of Fig.~\ref{comp_lada} shows the results for 
the Bondi-type accretion, i.e. $\dot{M} \propto M$ as 
emphasized in section~\ref{bondi-type}. The thick solid line 
reproduces the solid line of the upper panel to
facilitate 
comparison. Clearly the agreement 
is not as good as for the turbulent-type accretion 
for which $\dot{M} \propto M^{0.75}$, which is clearly 
in favor of this latter case. 
Therefore 
we restrict ourselves to this latter case in the following. 
Physically, 
it suggests that indeed large scale turbulent 
fluctuations may be regulating the accretion onto 
the proto-stellar clusters.

To summarize, the good agreement we obtain between our theory of 
proto-cluster formation and the data of embedded clusters 
suggests that $i)$ accretion-driven turbulence 
(Klessen \& Hennebelle 2010, Goldbaum et al. 2011) is 
 at play in low-mass proto-clusters, $ii)$ the accretion rate onto 
proto-clusters is reasonably described by Eq.~(\ref{accretion_rate})
with a value of $\alpha_{*,c} \simeq 2$ and 
$\dot{M} \propto M^{0.75}$. Therefore 
we adopt these values as fiducial parameters
for the remainder of the paper.

\section{Thermal balance and  mass distribution}
In this section we compute the temperature of the gas 
within the cluster and we also recall the 
principle and the expression of the Hennebelle \& Chabrier
theory considering a general equation of state. 

\subsection{Heating rate}
\subsubsection{Heating by turbulent dissipation}
As discussed in section~\ref{balance}, the turbulent 
energy within the cluster is continuously maintained 
by the accretion energy. The turbulent energy 
eventually dissipates, being converted into thermal energy, 
which is then radiated away. The gas within the 
cluster is thus subject to a mechanical heating  equal 
to the expression stated by Eq.~(\ref{energy_ext}). The mechanical 
heating per particle in the cluster is  given by 
\begin{eqnarray}
{\Gamma}_{\rm turb} = {6 \over 5} {G \dot{M} m_p \over R_*} {\rm erg \, s}^{-1}.
\label{mech_heat}
\end{eqnarray}
Note that as for the turbulent energy balance, a 
complete heat equation could be written, but 
as discussed above, large uncertainties hampered 
energy dissipation. 
Moreover, this heating represents an average quantity but may greatly 
vary through space and time in particular because turbulence 
is intermittent. Indeed, one may wonder whether the amount 
of energy dissipated per unit of time could not depend on the 
gas density.  Various authors (Kritsuk et al. 2007, 
Federrath et al. 2010) found a weak dependence of the Mach number 
on the density, the former decreasing as the latter increases. However, 
the dependence is extremely weak, ${\cal M} \propto \rho^{-0.05}$. 
It seems therefore a reasonable assumption to treat the mechanical 
heating as being uniform throughout the cluster.

With the help of the results of the preceding section, we can estimate 
this heating. For a cluster of radius $\simeq 1$ pc, the mass 
is about $10^3$ M$_\odot$  and the accretion rate 
$(2 \times 10^3/10^4) ^{0.75} \times 10^{-3} $ M$_\odot$ yr$^{-1}$ 
where $\alpha_{*,c}=2$ has been assumed.
Thus  ${\Gamma}_{\rm turb} \simeq 1.6 \times 10^{-27}$ erg s$^{-1}$.

\subsubsection{Cosmic ray heating}
In this work, we assume that the proto-cluster is embedded
in its parent clump, whose mass is a few times larger.
Since the total gas mass of the proto-cluster is  100
to 10$^4$ M$_\odot$, the mass of the parent clump is typically 
a few times this value, as discussed above. 
Hence the 
 column density of the parent clump is about
\begin{eqnarray}
\nonumber
R_c &=& \left(  { \alpha_{*,c}  M_*  \over  (4 \pi / 3)
  n_0 m_p (1 {\rm pc})^{0.7}} \right)^{1/2.3}, \\
\Sigma_c &=& 2 R_c n_c = 2 n_0 (1 {\rm pc})^{0.7} R_c^{0.3} \simeq 5-10 \times 10^{21} \, {\rm cm}^{-2}.
\label{dens_col_clump}
\end{eqnarray}
This implies that the visual extinction
 of the gas surrounding the proto-cluster is typically a few $A_v$. In the 
same way, it is typically about 10 or more throughout the cluster, as 
stated by Eq.~(\ref{cluster_rel}). 
This means that it is fair to consider that proto-clusters are sufficiently
embedded and optically thick to neglect the external UV heating. 
On the other-hand, cosmic rays are able to penetrate  even into 
well shielded clouds, providing a heating rate of
\begin{eqnarray}
{\Gamma}_{\rm cosmic} = 10^{-27} \left( {\zeta  \over 3 \times 10^{-17} {\rm s^{-1}}} \right){\rm erg \, s^{-1}},
\label{cosmic_heat}
\end{eqnarray}
where $\zeta$ is the mean cosmic ray ionization (e.g. Goldsmith 2001).
Note that our notation here differs from the choice that is often 
made because $\Gamma$ is the  heating per gas particle and not a 
volumetric heating.

Comparison between this value and ${\Gamma}_{\rm turb}$ reveals
that they are indeed comparable, with the higher dominating
 the former in massive proto-clusters.

\subsection{Cooling rate}
In the proto-cluster, the mean density is typically a few 1000 cm$^{-3}$
and the visual extinction is about 10. In these conditions, the 
gas is entirely molecular and well screened from  the UV, as 
already discussed. Two types
of cooling processes must be considered, the molecular line cooling 
and the  cooling by dust. 

\subsubsection{Molecular cooling}
For the molecular  cooling we use the tabulated values calculated 
and kindly provided by Neufeld et al. (1995). These calculations
assume that the gas is entirely screened from the UV background,
which is the case here. It is also assumed that the linewidth is the 
same for all species and is dominated by microturbulence. 
The corresponding values of the resulting cooling are displayed 
in Figure 3a-3d of Neufeld et al. (1995).
The table provided has temperatures between 10 and 3000 K, 
densities between 1000 and $10^{10}$ cm$^{-3}$ 
  with logarithmic increments. The 
column density per km~s$^{-1}$ is equal to 
$10^{20}$, $10^{21}$ or $10^{22}$ cm$^{-2}$.
The cooling per particles can typically be written as 
\begin{eqnarray}
{\Lambda_{\rm mol} \over n} = \Lambda_0 n^c T ^d \, {\rm erg \, s^{-1}},
\label{cooling_ana}
\end{eqnarray}
where $c$ is  small and $d$ about $2-2.5$
(see also Goldsmith 2001 and Juvela et al. 2001). 
Note that $\Lambda / n$ represents
the cooling per particle.

\subsubsection{Dust cooling}

The dust must also be taken into account in the thermal balance of the 
gas. The amount of energy exchanged per unit of time 
between a  gas particle and the dust is (e.g. Burke \& Hollenbach 1983, Goldsmith 2001)
\begin{eqnarray}
{\Lambda_{\rm dust} \over n} = 2 \times 10^{-33} n (T-T_d) 
\sqrt{{T \over 10 K}}  \, {\rm erg \, s^{-1}},
\label{cooling_dust}
\end{eqnarray}
where $T_d$ is the dust temperature.

\subsubsection{Dust temperature}

As shown by Eq.~(\ref{cooling_dust}), it is 
 necessary to know the dust temperature to compute $\Lambda_{\rm dust}$.
For this purpose we  closely follow the work of Zucconi et al. (2001). 
The dust temperature is the result of a balance between the dust emission 
and the absorption by the dust of the external infrared radiation, leading to 
\begin{eqnarray}
\int_0^\infty Q_\nu B_\nu(T_d(r)) d \nu = \int_0^\infty Q_\nu J_\nu(r) d \nu, 
\label{bal_dust}
\end{eqnarray}
where $Q_\nu$ is the grain absorption coefficient, $B_\nu$ is the Planck function
and $J_\nu(r)$ is the incident radiation field given by
\begin{eqnarray}
J_\nu(r) =  {J_\nu ^{is} \over 4 \pi} \int \exp( -\tau_\nu(r,\theta,\phi)) d \Omega, 
\label{J_ext}
\end{eqnarray}
where $J_\nu^{is}$ is the interstellar radiation and $\tau_\nu$ the 
optical depth.
Equation~(\ref{J_ext}) represents the  interstellar radiation attenuated
by the dust distribution within the proto-cluster and the parent clump.

The values of $Q_\nu$ and $J_\nu ^{is}$ are given in  Appendix B of Zucconi 
et al. (2001) and are  used here. To calculate $J_\nu(r)$
it is necessary to specify the spatial distribution of the dust. We 
assume that the proto-cluster is embedded into a spherical clump 
of mass $M_c =   \alpha_{*,c} M_*$, the  column density $\Sigma_c=2 R_c n_c$, 
can be estimated as in Eq.~(\ref{dens_col_clump}). 
We make the simplifying assumption that the external radiation 
that reaches the edge of the proto-cluster  is
$J_\nu(0) = J_\nu^{is} \exp(-\Sigma_c/2 \times Q_v)$. Hence because the cluster is assumed to be 
on average uniform in density, the radiation field within the cluster can be 
estimated as described in Appendix A of Zucconi et al. (2001), that is 
\begin{eqnarray}
\nonumber
J_\nu(r) = {J_\nu(0) \over 4 \pi} \int _0^\pi \exp(-\tau_\nu(r,\theta)) 2 \pi \sin(\theta) d\theta, \\
\tau_\nu(r,\theta) = Q_\nu \Sigma _*^c (\sqrt{1-(r/R_*)^2 \sin^2(\theta)}-(r/R_*) \cos(\theta)),
\label{J_spa}
\end{eqnarray} 
where $\Sigma_*^c$ is the proto-cluster column density toward the center. 
Equation~(\ref{J_spa})
  represents the integration of the radiative transfer equation in all directions
through a cloud of radius $R_*$ that has a uniform density.

To find the dust temperature at a given position within the proto-cluster, we 
solve  Eq.~(\ref{bal_dust}) using the wavelength dependence 
 of $J_\nu^{is}$ and $Q_\nu$
given in  Appendix B of Zucconi et al. These 
values represent a fit from the interstellar radiation field given by Black (1994) and 
grain opacities from Ossenkopf \& Henning (1994), respectively.
For the latter, standard grain abundances are assumed.
In practice this entails the calculations of integrals of the type 
indicated in Eq.~(9) of Zucconi et al. Because we do not only calculate the 
dust temperature at the proto-cluster centre, Eq.~(10) of Zucconi et al. 
must be replaced by $\int _{\lambda_{min}} ^{\lambda_{max}} t^q / (\exp(t/\beta_i)-1)
J_t (r) dt$. Since $J_t$ needs itself the calculation of an integral (that we express
as Eq.~(A.3) of Zucconi et al.) as shown 
by Eq.~(\ref{J_spa}), the calculations require some integration time. 
We verify that our results compare well to the results shown in Fig.~2 of 
Zucconi et al. and we  obtain temperatures of 
about $\simeq 8$~K in the 
proto-cluster center and 9.5 K at the edge. 

Finally, because we ought to determine a single dust temperature within the 
proto-cluster, we compute the mean dust temperature 
as $\bar{T}_d = \int _0^{R_*} T_d (r) 4 \pi r^2 dr / (4 \pi/3 R_*^3)$, which 
is typically equal to about 9 K.

\subsection{Temperature distributions}

To find the temperature of a proto-cluster whose density and column 
density per km~s$^{-1}$ are known, we must solve for the thermal equilibrium
\begin{eqnarray}
n ({\Gamma}_{\rm turb} + {\Gamma}_{\rm cosmic}) = 
\Lambda_{\rm mol} + \Lambda_{\rm dust}.
\label{therm_bal}
\end{eqnarray}

We start by computing $T_{dust}$ as described above. To solve Eq.~(\ref{therm_bal}), we use an iterative method,
 performing a first order interpolation 
in logarithmic density, column density per km~s$^{-1}$ and temperature to derive
 the tabulated molecular cooling. 
Note in passing that the column density per km~s$^{-1}$, which 
is equal to $\Sigma_* / (\sigma_* / \sqrt{3} )$ 
is decreasing with the proto-cluster mass since 
$\Sigma_*$ is constant but $\sigma \propto R_*^{1/2}$. Thus 
the more massive clusters can cool more efficiently. On the other 
hand their heating is also more intense because
 it is proportional to $\dot{M}$,
which scales as $M_*^{0.75}$. 

\subsubsection{Mean temperature and Jeans mass}

Figure~\ref{temp_mass} portrays the mean gas temperature as a 
function of proto-cluster mass for $\alpha_{*,c}=1$, 2 and 4. 
Evidently, the temperature
increases from about 10 K to almost 20 K 
for the biggest mass considered.
Since the gas density is decreasing as the proto-cluster mass is increasing,
this implies that the Jeans mass is unavoidably increasing with 
the proto-cluster mass as displayed in Fig.~\ref{jeans_mass}.
Indeed, it shows that as expected the Jeans mass within the proto-clusters
increases from about 5 to $\simeq$25 $M_\odot$ as $M_*$ evolves
from 10$^2$ to 10$^4$ $M_\odot$. This could suggest that, indeed, 
the IMF could vary within clusters. However, as we will see below
this is not the case. The primary reason is that the peak of the IMF 
also depends $i)$ on the Mach number and $ii)$ on the equation of state,
i.e. the temperature dependence on the density.

\begin{figure} 
\includegraphics[width=9cm]{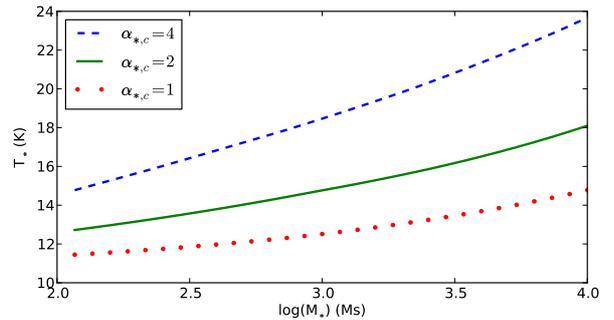}
\caption{Mean gas temperature  for various proto-cluster masses.}
\label{temp_mass}
\end{figure}

\begin{figure} 
\includegraphics[width=9cm]{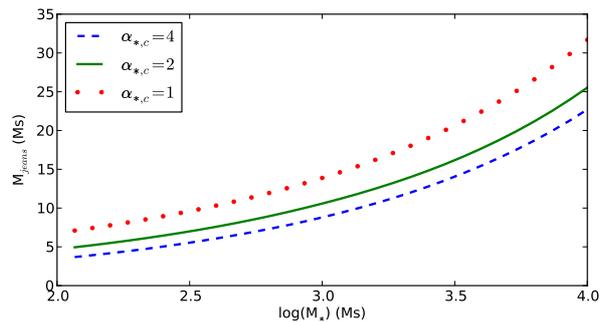}
\caption{Jeans mass vs proto-cluster masses.}
\label{jeans_mass}
\end{figure}

\subsubsection{Temperature distribution within proto-clusters}
\label{tempe_dist}

As discussed in Hennebelle \& Chabrier (2009, HC2009), the equation of state
has a significant influence on the core mass function. Knowing 
the mean temperature only, does not appear to be sufficient, therefore. Instead
one needs to know the complete equation of state, that is to say, 
how the temperature varies with the density within the proto-clusters. 
For that purpose, we  compute the temperature at various densities,
assuming that the turbulent heating $\Gamma_{\rm turb}$ and 
the gradient per km s$^{-1}$ used for the molecular 
cooling correspond to the mean  conditions within proto-clusters 
of mass $M_*$. This means 
that we are assuming that both the heating and the cooling, only 
depend on the large scale conditions and do not vary locally. 

Figure~\ref{dens_temp} shows the gas temperature as a function of 
density for various proto-cluster masses.
For densities between $10^3$ and $\simeq 10^6$ cm$^{-3}$, 
the temperature decreases from about 20 to $\simeq 8$K, leading 
to an effective adiabatic exponent of about $\simeq 0.85-0.9$.
 At higher densities, the temperature increases slightly due to the influence
of the dust. The minimum temperature is reached for 
$n \simeq 5-8 \times 10^5$ cm$^{-3}$. This compares reasonably well 
with the temperature 
distributions displayed in Fig.~2 of Larson (1985).
The proto-cluster mass does not have a drastic influence on the temperature
distribution with variations of a few degrees only.

\begin{figure} 
\includegraphics[width=9cm]{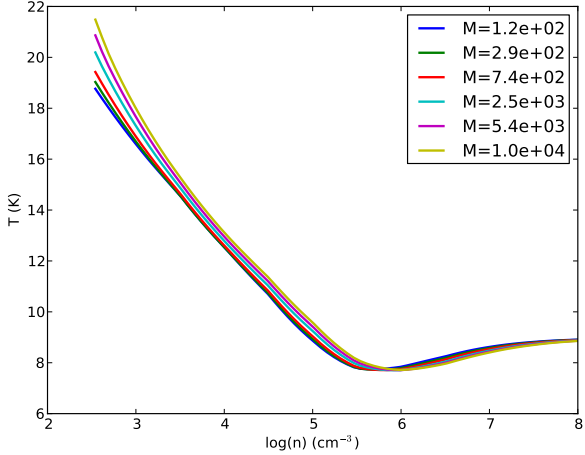}
\caption{Temperature vs density for various proto-cluster masses.}
\label{dens_temp}
\end{figure}

\subsection{Mass distribution: Hennebelle-Chabrier theory}
In this section, we briefly describe the ideas and the formalism
of the Hennebelle-Chabrier theory of  star formation 
(HC2008, HC2009, HC2011) that is used 
in this manuscript to infer the IMF within clusters. 
In this theory the prestellar cores that eventually lead
to protostars and then stars correspond to the 
turbulent density fluctuations that arise in a supersonic 
medium,  which become sufficiently dense to be self-gravitating. 
These fluctuations are counted by using the formalism 
developed in cosmology by Press \& Schechter (1974) although 
with a  formulation close to the approach of Jedamzik (1995).
Recently, Hopkins (2012) formulates the problem using the 
excursion set theory (e.g. Bond et al. 1991) formalism. He also 
extends the calculation by considering the whole galactic disk. 
His results regarding the small scale self-gravitating fluctuations 
(the last crossing barrier) are undistinguishable from the HC2008
result at small masses and only slightly different at high mass.

According to various simulations  of hydrodynamic or MHD supersonic 
turbulence, the density PDF is well 
represented  in both cases by a lognormal form,
\begin{eqnarray}
\label{Pr0}
{\cal P}(\delta) &=& {1 \over \sqrt{2 \pi \sigma_0^2}} 
\exp\left(- { (\delta - \bar{\delta})^2 \over 2 \sigma_0 ^2} \right) , \;
 \delta = \ln (\rho/ \rho_0 ), \\
 \bar{\delta}&=&-\sigma_0^2/2 \;
,  \; \sigma_0^2=\ln (1 + b^2 {\cal M}^2),
\nonumber
\end{eqnarray}
where ${\cal M}$ is the Mach number, $b \simeq 0.5-1$
(Kritsuk et al. 2007, Audit \& Hennebelle 2010, Federrath et al. 2010)
and $\rho_0$ is the mean density.

The self-gravitating fluctuations are  
determined by 
identifying  the structures of mass $M$
in the cloud's random field of density 
fluctuations. These  are gravitationally 
unstable at scale $R$, according 
to the virial theorem. This condition defines a {\it scale-dependent} 
(log)-density threshold,
$\delta_R^c=\ln(\rho_c(R)/{ \rho_0})$, or equivalently, a scale-dependent
 Jeans mass, $M_R^c$
\begin{eqnarray} 
M_R^c = a_J^{2/3} 
\left( {  (C_s)^2 \over G    } R + {V_0^2  \over 3\, G  } 
\left({R \over 1 {\rm pc}}\right)^{2\eta} R \right),  
\label{crit_Mtot}
\end{eqnarray}  
where $C_s$ is the sound speed, G the gravitational constant, $a_J$ 
a constant on order of unity 
while $V_0$ and $\eta$ determine the rms velocity,
\begin{eqnarray}
\langle V_{\rm rms}^2\rangle =  V_0^2 \times \left( {R \over  1 {\rm pc}} 
\right) ^{2 \eta}.
\label{rms_vel}
 \end{eqnarray} 
 Because a
 fluctuation of scale $R$ is replenished within a typical crossing time
$\tau _R$,  and is therefore  
replenished a number of time equal to $\tau ^0 _{ff} / (\phi_t \tau_{R,ff})$, 
where
 $\tau_{R,ff}=\tau_R/\phi_t$ is the freefall time at scale $R$  
 (see Appendix of HC2011), 
 i.e. at density $\rho_R\sim M_R/(4 \pi / 3 R^3)$, 
HC2011  
includes this condition into the formalism originally 
developed in HC2008.
This yields
for the number-density mass spectrum of gravitationally bound structures, 
${\cal N} (M)=d(N/V)/dM$
\begin{eqnarray}
\label{n_general}
{\cal N} ( M_R)  \simeq
 { {\rho_0} \over M_R} 
{dR \over dM_R} \times \left( -{d \delta_R  \over dR} e^{\delta_R} 
({\tau ^0 _{ff} \over \tau_R}) {\cal P}( \delta_R) 
 \right),
\label{spec_mass}
\end{eqnarray}
which is  Eq.~(6) of HC2011 and 
except for the time ratio ${\tau ^0 _{ff} \over \tau_R}$, 
is similar to 
Eq.~(33) of HC2008. Note that we have dropped 
the second term which appears in Eq.~(33) and 
which entails the derivative of the density 
PDF. For the strongly self-gravitating
 regimes that we are considering here, this 
approximation is well satisfied. 

Equation~(\ref{n_general}) gives the mass
spectrum that we are computing in the following. It depends 
on $M_R$ and $d M_R / d R$, which can be obtained from 
Eq.~(\ref{crit_Mtot}). To proceed, it is more convenient to 
normalize the expressions.
After normalisation, Eq.~(\ref{crit_Mtot}) becomes
\begin{eqnarray}
\nonumber
\widetilde{M}_R^c =  M / M_J^0 = \widetilde{R}\,
\left( f(\rho) +  {\cal M}^2_* \widetilde{R}^{2 \eta}\right),\\
 f(\rho) =f( {\tilde{M_R^c} \over \tilde{R}^3}) = 
{ C_s^2 \over  (C_s^0)^2},
\label{mass_rad}
\end{eqnarray}
where $C_s^0=C_s(\rho_0)$, $M_J^0$, $\lambda_J^0$ and ${\cal M}_*$ are given by
\begin{eqnarray}
\nonumber
M_J^0&=& a_J\,{ (C_s^0)^3 \over \sqrt{G^3 \rho_0}}\approx 1.0\,\, a_J \,({T \over 10\,{\rm K}})^{3/2}\,
({\mu \over 2.33})^{-1/2}\, \\ &\times&({ n_0 \over 10^4\,{\rm cm}^{-3}})^{-1/2}\, {\rm M}_\odot ,
\label{mjeans} \\
\nonumber
\lambda_J^0&=& \left( {a_J \over C_m} \right)^{1/3}{(C_s^0)\over \sqrt {G \rho_0}}\approx 0.1\, a_J^{1/3} \,({T \over 10\,{\rm K}})^{1/2}\,
({\mu \over 2.33})^{-1/2}\, \\ &\times& ({n_0 
\over 10^4\,{\rm cm}^{-3}})^{-1/2}\,\, {\rm pc},
\label{ljeans}
\end{eqnarray}
\begin{eqnarray}
\nonumber
{\cal M}_* &=& { 1  \over \sqrt{3} } { V_0  \over C_s}\left({\lambda_J^0 \over  
 1 {\rm pc} }\right) ^{ \eta} \\
&\simeq& (0.8-1.0) \,\left({\lambda_J^0\over 0.1\,{\rm pc}}\right)^{\eta}\,\left({C_s\over 0.2 \, {\rm km \, s}^{-1}}
\right)^{-1}, 
\label{mass_star}
\end{eqnarray}
$a_J$ and $C_m$ being  dimensionless geometrical factors on the order of unity. 
Taking for example the standard definition of 
the Jeans mass, as the mass enclosed in a sphere of diameter equal to the Jeans length, we get 
$a_J=\pi^{5/2}/6$ while $C_m=4 \pi / 3$ (e.g. HC2009).

With Eqs.~(\ref{Pr0}) and~(\ref{n_general}), we obtain 
\begin{eqnarray}
\nonumber
{\cal N}(\widetilde{M}) &=& -{1 \over \phi_{t}} 
{ \rho_0 \over M_J^{0} \widetilde{M}}
 \left( { \widetilde{M}_R^c \over \widetilde{R} ^3 } \right)^{1/2}
 {d \widetilde{R} \over d \widetilde{M}_R^c}  {d \delta_R^c\over d\widetilde{R}}  \\
&\times& {1 \over \sqrt{2 \pi \sigma^2} } \exp \left( -{(\delta_R^c)^2 \over 2 \sigma^2} + 
{\delta _R^c \over 2} - {\sigma^2 \over 8}  \right).
\label{mass_spec}
\end{eqnarray}

An important difference to the work of HC2009
is that the equation of state shown in Fig.~\ref{dens_temp} is not polytropic, 
but fully general. 
To obtain $d \tilde{M}_R^c / d \tilde{R}$, we must  differentiate 
Eq.~(\ref{mass_rad}), which leads to (see Appendix of  HC2009)
\begin{eqnarray}
{ d \tilde{M}_R^c \over d \tilde{R}}  =
{ f(  {\tilde{M_R^c} \over \tilde{R}^3} )   - 3 {\tilde{M_R^c} \over \tilde{R}^3} f'({\tilde{M_R^c} \over \tilde{R}^3}) + (2 \eta +1) {\cal M}_*^2 \tilde{R}^{2 \eta} \over  1 - { 1 \over \tilde{R}^2 } f'(  {\tilde{M_R^c} \over \tilde{R}^3} ) }.
\label{der_eos}
\end{eqnarray}
With this last equation, all quantities appearing in Eq.~(\ref{mass_spec})
are known and the mass spectrum of the condensations can be computed.
Note that $M_R$ must be computed numerically 
from Eq.~(\ref{mass_rad}).

\section{Results: mass distribution of self-gravitating condensations
 in clusters}
In this section we calculate the mass spectrum of the 
self-gravitating fluctuations and  discuss its dependence 
on the various parameters. 

\subsection{Preliminary considerations}
Before presenting the complete distribution, we start by discussing 
the position dependence of the distribution peak. For that 
purpose we make  various simplifying assumptions.

\subsubsection{Peak position: dependence on the Jeans mass and Mach number}
Unlike what is often assumed in the literature, it is unlikely that the 
peak of the CMF/IMF solely depends on the Jeans mass. 
In particular, it likely depends on the Mach number because compressible
turbulence creates high density regions where the Jeans mass 
is smaller. Indeed, in any turbulent medium, there is a distribution 
of Jeans masses rather than a single well-defined value. 
The peak position has been calculated by HC2008
who  show (their Eq.~46 ) that 
$M_{\rm peak} = M_J^0/(1+b^2 {\cal M}^2)^{3/4}$ (note that 
in HC2008 $b$ stands for $b^2$ as different notations
were used). However, 
as discussed above, time-dependence is not considered in 
HC2008 while it is taken into account in 
Eq.~(\ref{mass_spec}) through the term $1/\tau_{R}$. 
To calculate the peak position in this case, we proceed as in 
HC2008, that is to say we neglect the turbulent
support 
(i.e. we set ${\cal M_*}=0$), which has little influence on the 
peak position. We also assume strict isothermality within the cluster, 
that is to say, we set $f=1$. As discussed in HC2009,
the equation of state indeed has an influence on the peak position. 
However, as portrayed in Fig.~\ref{dens_temp}, the deviation 
from the isothermal case is not very important with an effective adiabatic 
index of 0.85-0.9. Moreover, our goal here is more to discuss the dependence qualitatively  
rather than getting an accurate estimate, which is calulated later 
in the manuscript.

Assuming ${\cal M_*}=0$ and $f=1$, it is easy to take the 
derivative of Eq.~(\ref{mass_spec}) and to show that the maximum 
of ${\cal N}(\tilde{M})$
is reached for $\tilde{M}=1/(1+b^2 {\cal M}^2)$,
 which implies 
that the distribution peak is reached at 
\begin{eqnarray}
M_{\rm peak} = { M_J^0 \over 1+b^2 {\cal M}^2 }.
\label{peak}
\end{eqnarray}
The difference to the expression presented in HC2008
comes from the fact that more small structures form when time dependence
is taken into account since the small scale fluctuations are rejuvenated many times
 while the larger ones are still evolving.

Note that as already stressed, the peak position does not depend on the 
Jeans mass only, but also on the Mach number. For a typical $b^2$ of the order
of $\simeq$0.5 (see Federrath et al. 2010) and a Mach number of about 
5, we derive that the peak position is typically shifted by a factor of about 
10 with respect to the mean Jeans mass.

\subsubsection{Dependence of the peak position on the cluster mass}
To estimate the peak position, we must therefore estimate the Jeans mass and 
the Mach number as a function of the cluster parameters.
Since the Jeans mass is equal to $(\pi^{5/2}/6) C_s^3 / \sqrt{\rho_0 G}$, we obtain
that 
\begin{eqnarray}
M_{\rm peak} \simeq {\pi^{5/2} \over 6} {C_s^5 \over b^2 G^{1/2}  
(\rho_*)^{1/2}\sigma_*^2 }.
\label{peak2}
\end{eqnarray}

To estimate the sound speed we must compute the mean temperature 
within the proto-cluster. For the sake of simplicity, we 
 consider in this analytical estimate the turbulent heating $\Gamma _{turb}$ 
and the molecular cooling $\Lambda _{mol}$ 
given  by Eqs.~(\ref{mech_heat}) and~(\ref{cooling_ana}), respectively.
Writing $\Gamma _{turb} = \Lambda _{mol} / n$, we obtain the 
temperature and thus the sound speed
\begin{eqnarray}
\nonumber
C_s^5 &=& \left( { k \over m_p} \right)^{5/2}
\left( {6 \over 5} {G m_p  \over  1 \, {\rm pc}}
 \right)^{5/(2d)}
\left( {Q_0 (M_*^0)^{0.75} \over   \Lambda_0 n^c }  \right)^{5  / (2 d)} 
\\
&\times& \left( { R_* \over 1 \, {\rm pc}} \right)^{5 / (4d)}.
\label{sound}
\end{eqnarray}
On the other hand, with Eqs.~(\ref{cluster_rel})-(\ref{cluster_sig}),
we derive
\begin{eqnarray}
{1 \over \sqrt{\rho_*} \sigma^2} = 
 { \sqrt{4 \pi / 3 (1 \, {\rm pc})^3 } \over  
(2 \sqrt{3} G)^{2/3}   }
   {  1 \over Q_0^{2/3}  (M_*^0)^{1/2} } 
\left( {R_* \over 1 {\rm pc}} \right) ^{-1/2}.
\end{eqnarray}
This leads to 
\begin{eqnarray}
\nonumber
M_{\rm peak} &\simeq& 
K  { Q_0^{5  / (2 d)-2/3} (M_*^0)^{15  / (8 d)-1/2}  \over \Lambda_0 ^{5  / (2 d)} n_c^{5c/(2d)} } 
\left( { R_* \over 1 \, pc} \right)^{5 / (4d) - 1/2}, \\
\nonumber
K &=& {\pi^{5/2} \over 6} {1 \over G^{1/2} b^2} \left( { k \over m_p} \right)^{5/2}
\left( {6 \over 5} {G m_p \over  1 \, {\rm pc} }
 \right)^{5/(2d)} \\
&\times& { \sqrt{4 \pi / 3 (1 \, {\rm pc})^3 } \over  
(2 \sqrt{3} G)^{2/3} }.
\label{peak2}
\end{eqnarray}
As already mentioned, typical values of  $d$ are 2-2.5 while 
$c$ is typically low and close to zero. Ignoring the dependence
on $n^c$, the peak position consequently 
has a weak dependence on the cluster radius, which typically depends on $R$ as $R_*^{1/8-0}$. 
This is the result of partial compensation of
 the sound 
speed, the density and the velocity dispersion.
Thus we can conclude  that in the regime where 
turbulent heating dominates over the cosmic ray heating, 
the initial mass function is expected to weakly depend 
on the cluster masses.  According to our estimate, 
this is the case for large mass clusters, i.e. 
with masses
above $\simeq 10^3 M _\odot$. Note that it should be kept in mind that
in this analysis, isothermality is 
assumed, which is also a simplification.

Before solving for the whole mass spectrum obtained
with a complete equation of state, it is worth computing
the dependence of $M_{\rm peak} \simeq M_J^0/(1+b^2 {\cal M}^2)$
on the proto-cluster mass and on $Q_0 \propto \alpha_{*,c}^{0.75}$, the parent clump over proto-cluster 
 mass ratio using the complete 
thermal balance stated by Eq.~(\ref{therm_bal}) rather than the simplified
approach used above.
Figure~(\ref{pic_imf}) portrays the value of 
$M_{peak} = M_J^0/(1+b^2 {\cal M}^2)$ as a function 
of the proto-cluster mass for three values 
of $\alpha_{*,c}$ and $b=0.6$. This last value corresponds 
to a mixture of solenoidal and compressive modes
(Federrath et al. 2008).
The trends are similar to what has been 
inferred from Eq.~(\ref{therm_bal}), that is to say, the 
peak position  weakly depends on the proto-cluster mass
and more strongly on the accretion rate controlled by $\alpha_{*,c}$.
However, it is remarkable that for proto-clusters whose 
masses are between $10^2$ and $10^4$ M$_\odot$,
and within the ranges $\alpha_{*,c} = [1,4]$, which 
encompasses most of the data points reported in Fig.~\ref{comp_lada}, 
the peak  position varies by  a factor of only 3 
and only 2 for $M_c$ between 10$^3$ and 10$^4$ M$_\odot$. 

\begin{figure} 
\includegraphics[width=9cm]{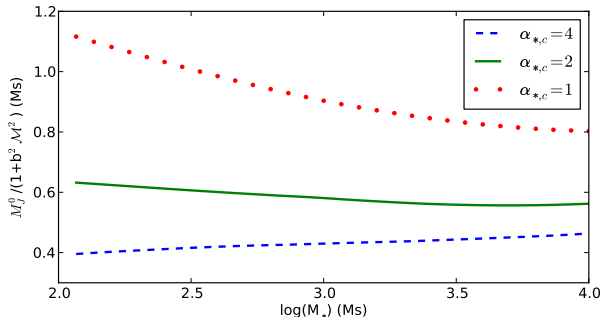}
\caption{Expected position of the CMF peak 
as a function of the proto-cluster mass for
3 values of the  ratio of parent clump
over proto-cluster mass ratio, $\alpha_{*,c}$.}
\label{pic_imf}
\end{figure}

\subsection{Fiducial parameters}
We now present the complete mass spectrum obtained by 
iteratively solving 
Eq.~(\ref{mass_rad}) and computing Eqs.~(\ref{mass_spec})
and~(\ref{der_eos}) using the temperature distribution 
obtained in section~\ref{tempe_dist} and portrayed in 
Fig.~\ref{dens_temp}. Because the latter is 
not analytical, it is difficult to take its derivative as
required by Eq.~(\ref{der_eos}). Thus for this purpose we  
 first obtain a fit of the temperature using a high order polynomial. 

\setlength{\unitlength}{1cm}
\begin{figure} 
\begin{picture} (0,14)
\put(0,7){\includegraphics[width=9cm]{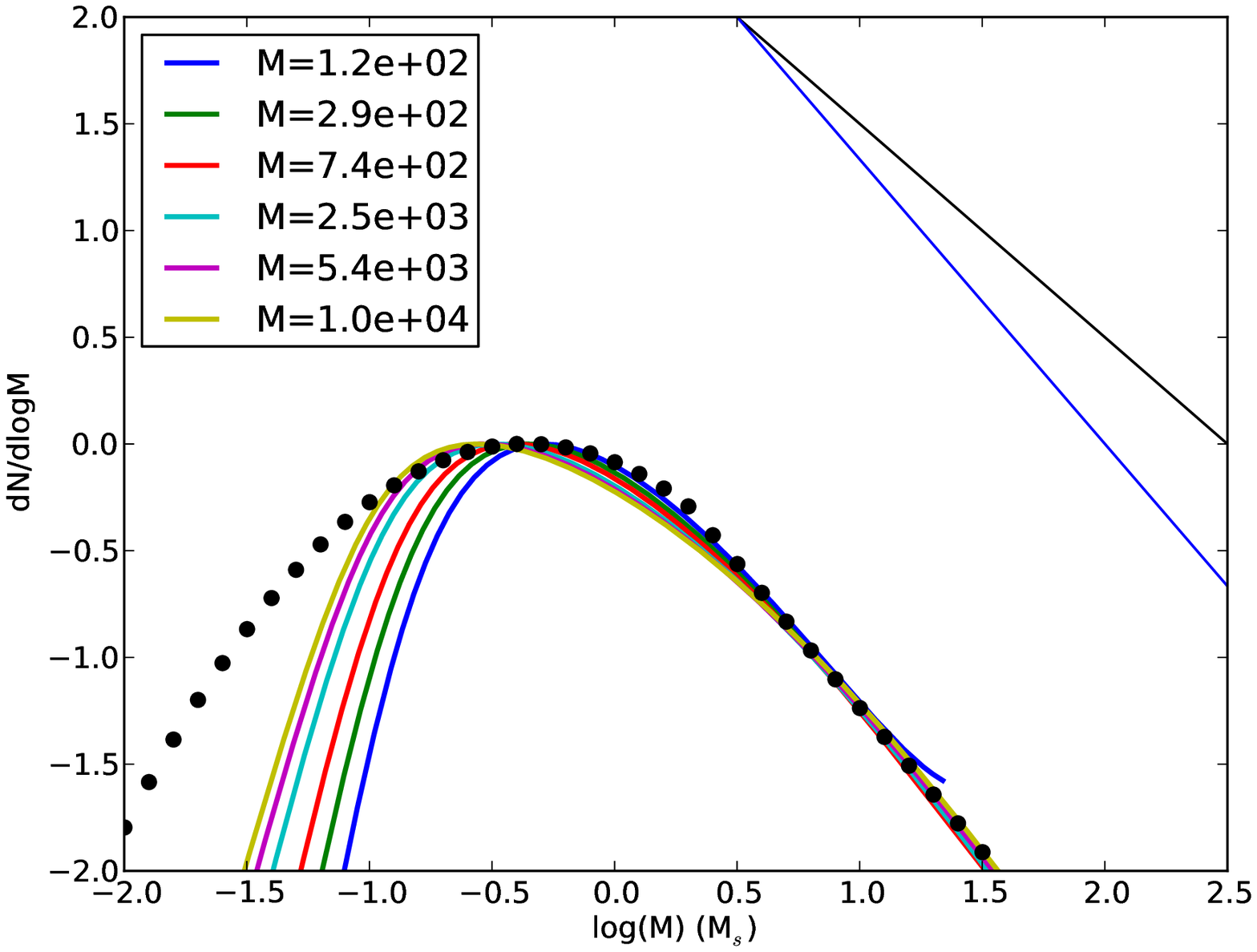}}
\put(0,0){\includegraphics[width=9cm]{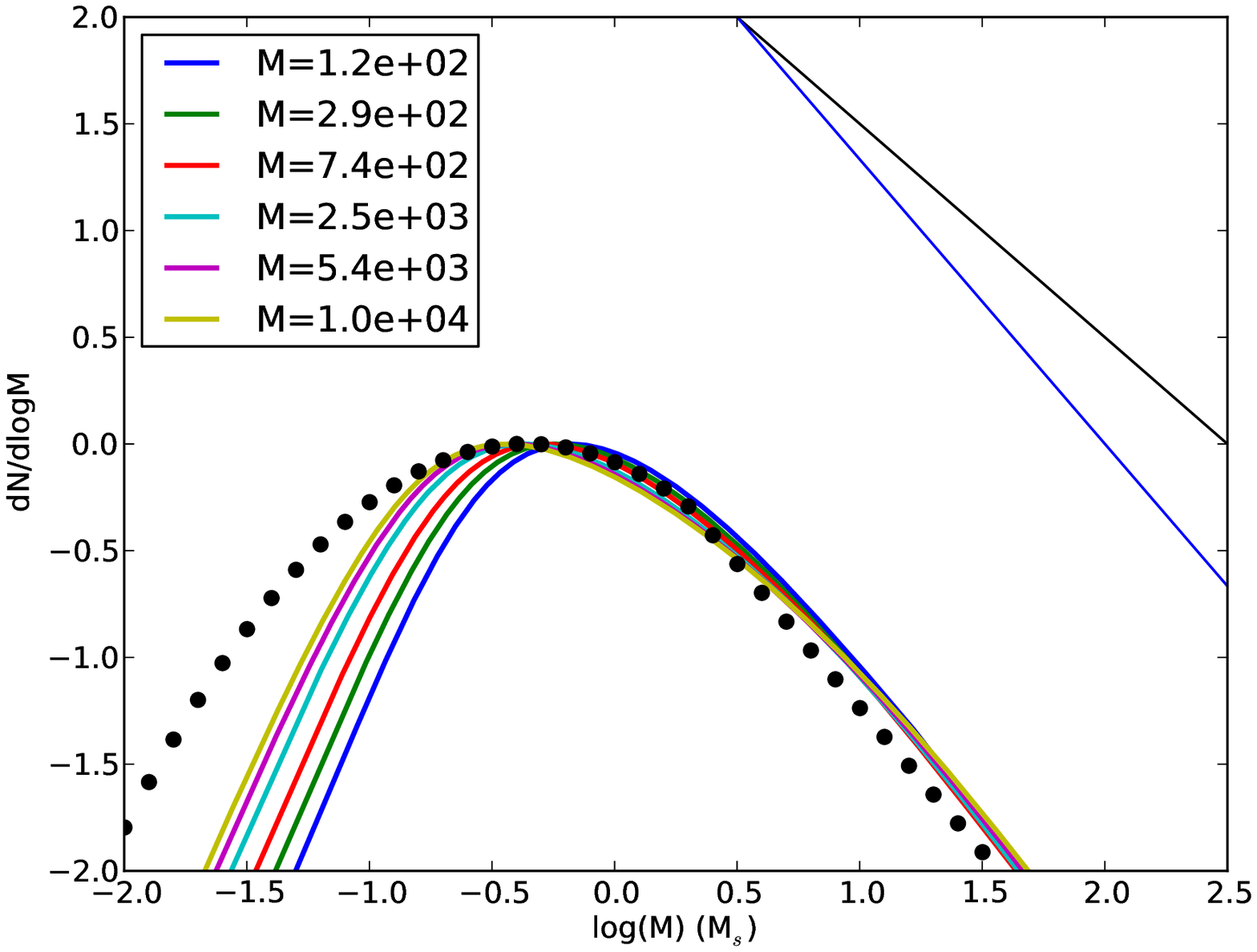}}
\end{picture}
\caption{Mass spectra for various cluster masses
and the fiducial value $\alpha_{*,c}=2$.
The dots represents the Chabrier IMF. The two lines
represents the distribution 
$M^{-2}$ and $M^{-2.3}$. The upper panel displays the 
mass distribution obtained with Eq.~(\ref{mass_spec})
while the lower panel shows this mass distribution convolved 
by a Gaussian of dispersion $\sigma=M/2$. }
\label{dN_dlogM}
\end{figure}

Upper panel of Fig.~\ref{dN_dlogM} portrays the resulting 
mass spectra for various cluster masses as labeled 
in the figure. The dots represent the Chabrier
 IMF (Chabrier 2003)
shifted by a factor of about 2. 
Indeed, since we compute the mass function of self-gravitating 
condensations, this shift is necessary to account for 
 the observed shift between the 
CMF and the IMF (e.g. Alves et al. 2007, Andr\'e et al. 2010)
and for the theoretical estimate for the 
 core-to-star efficiency (Matzner \& McKee 2000, Ciardi \& Hennebelle 2010).
Because we do not consider
 any cluster mass distribution and  efficacity 
problem here, we  set the distribution 
maximum to 1. This allows us to compare the shape of 
the various distributions more easily.

Various points are worth discussing. 
First of all, the distributions  all 
 agree with the Salpeter and Chabrier IMF at high masses. 
As discussed in HC2008, this is because the high mass part 
is essentially due to the turbulent support, which 
does not change much in this regime. Moreover, the high mass 
part of the distribution tends to slowly depend 
on the parameters that control the turbulence, namely 
${\cal M}$ and ${\cal M}_*$.
Second, the peak position varies by about 
a factor 2 as the proto-cluster mass changes from 
$\simeq 10^2 M_\odot$
 to about $10^4 M_\odot$ while the distributions corresponding to 
$M_*=10^3$ and $10^4$ $M_\odot$ are nearly indistinguishable.
This weak variation is, as discussed in the previous section,
a consequence of various quantities compensating each other.
Third, the position of the peak itself is close to 
the peak of the Chabrier IMF shifted by a 
factor of 2.

\subsection{Link between the CMF and the IMF}

An important question regarding the mass distribution 
is the link between the IMF and the CMF, which is not 
expected to be as simple as a unique efficiency (e.g. Alves et al. 2007, 
Goodwin et al. 2008)
 on the order of 2-3. It is instead expected that the two distributions
 are correlated with some dispersion because the initial mass and 
 velocity distributions within the cores vary from one core to another,
therefore leading to different evolutions. In other words, the virial 
theorem is not representing the problem in its 
full complexity and the initial conditions and the boundary
conditions (which enter through the surface 
terms, see e.g. Dib et al. 
2007) influence the mass of the objects that form within the collapsing
cores. 
Indeed, Smith et al. (2008) 
have shown that in hydrodynamical simulations, the correlation 
between the mass of the sink particles and the cores in which 
they are embedded is very good during the first freefall times 
and becomes less tight after $\simeq 5$ freefall times.
This is because initially the sink particles are 
accreting the mass of the parent core. As time 
evolves, the material 
that falls onto the sinks comes for further away and is 
less and less correlated to the initial mass reservoir. 
More quantitatively, Chabrier \& Hennebelle (2010) have 
shown that in the simulations of Smith et al. (2008), 
the correlation between the CMF and the 
sink particles mass function, which is likely representing the
IMF, can reasonably be reproduced at intermediate time (say 3-5 freefall time)
by a Gaussian distribution with a width,
$\sigma$, on the order of $M/3-M/2$.
To take this into account, we have convolved the mass distribution 
portrayed in the upper panel of Fig.~\ref{dN_dlogM}
with a Gaussian distribution of width $\sigma_{conv}=M/2$.
\begin{eqnarray}
{dN  \over dM} _{conv} (M)  = 
{ \int _0 ^\infty   {dN  \over dM} (M')
\exp \left( { -(M-M')^2 \over 2 \sigma_{conv} ^2} \right) dM'
\over 
 \int _0 ^\infty 
\exp \left( { -(M-M')^2 \over 2 \sigma_{conv}^2} \right) dM'}.
\end{eqnarray}
Note that in principle, since the efficiency of the accretion 
is lower than one, the distribution should be shifted 
toward smaller masses (e.g. Alves et al. 2007, Andr\'e et al. 2010). 
In practice, to facilitate the comparison
with $dN/dM$, we have not shifted it here. 

As expected, the distribution $dN  / dM _{conv}$ is 
slightly broader than the distribution $dN/dM$ in particular 
at low masses whereas, the high mass 
part and the peak position are almost unchanged. This 
effect could therefore in particular contribute to the formation of 
low mass brown dwarfs. To clarify, because some cores
are marginally bound, for example because their velocity field
is initially globally diverging, 
few  objects significantly less massive than the parent core mass form, 
most of the envelope being then dispersed in the 
surrounding ISM. 

To summarize, for an accretion rate that allows one to reproduce
the mass-size relation of clusters, we self-consistently predict
a distribution of self-gravitating structures that 
$i)$ is very close to the field IMF inferred by Chabrier (2003), 
$ii)$ is almost independent on the cluster mass for 
gas masses larger than $10^3 M_\odot$. 
We stress that if more variability is expected for low mass
clusters, it is very difficult to infer any reliable 
statistics and thus no data are available in this regime.

\subsection{Dependence on the accretion rate}
Because the accretion rate onto the proto-clusters
of a given mass is likely varying, as suggested 
by Fig.~\ref{comp_lada}, we explore the 
effect of this parameter on the whole mass spectrum.
As anticipated in Fig.~\ref{pic_imf}, it has 
some influence on the peak position.
As discussed in section 2.1, we use the 
 parent clump mass over proto-cluster mass  ratio 
to quantify the accretion rate following 
Eqs.~(\ref{accretion_rate})-(\ref{accretion})
and we select the two values  $\alpha_{*,c}=1$ and 4, 
which as shown by  Fig.~\ref{comp_lada} encompass almost all  points of the observational distribution.

The mass distributions obtained for 
$\alpha_{*,c}=1$ and 4 are 
 displayed in upper and lower panels of 
 Fig.~\ref{dN_dlogM_1}, respectively,  which 
shows that the trends inferred in Fig.~\ref{pic_imf}
are well confirmed. The mass distribution is shifted 
toward larger masses for a lower accretion rate ($\alpha_{*,c}=1$)
and toward smaller masses for a higher  accretion rate ($\alpha_{*,c}=4$).
Apart from this, the general behaviour is almost identical to the 
fiducial case $\alpha_{*,c}=2$, that is to say, the 
distributions weakly depend on the proto-cluster gas masses
for $M_* \simeq 10^3-10^4$ M$_\odot$ and the high mass part 
is always close to the Salpeter IMF.


Importantly enough, we note that within the range $1< \alpha_{*,c} <4$
the variation of the 
mass spectrum with the accretion rate remains 
limited to a factor of about 2  almost everywhere,  especially for the proto-clusters,
whose mass of gas is between 10$^3$ and $10^4$ M$_\odot$.

\begin{figure} 
\begin{picture} (0,13)
\put(0,7){\includegraphics[width=9cm]{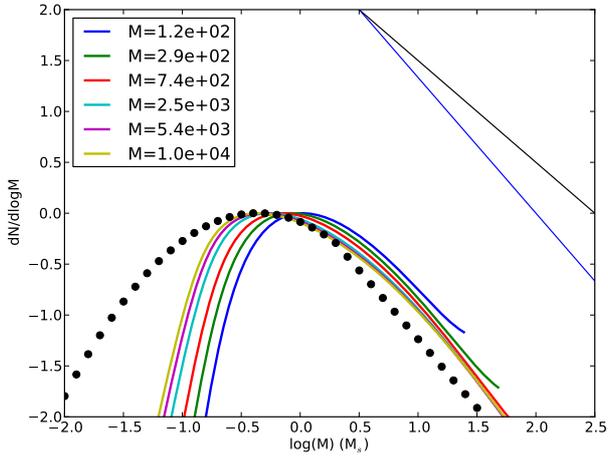}}
\put(0,0){\includegraphics[width=9cm]{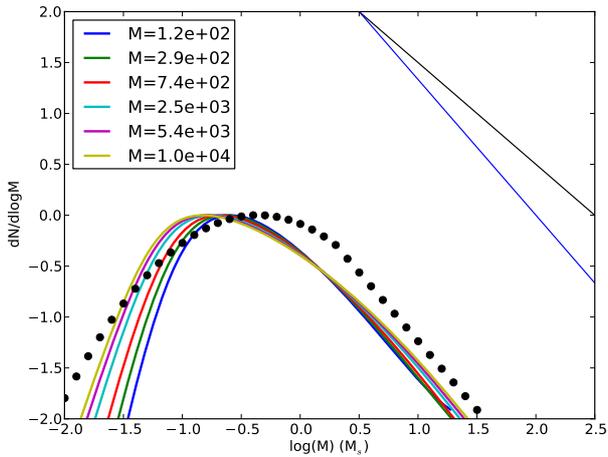}}
\end{picture}
\caption{
Mass spectra for various cluster masses,
upper panel  $\alpha_{*,c}=1$, lower panel  $\alpha_{*,c}=4$.
As in Fig.~\ref{dN_dlogM}
the dots represents the Chabrier IMF. The two lines
represents the distribution 
$M^{-2}$ and $M^{-2.3}$.}
\label{dN_dlogM_1}
\end{figure}


\subsection{Dependence on cosmic rays and radiation}
Another source of possible variability are the heating sources, 
namely cosmic rays and interstellar radiation
field, which 
heat the gas and the dust respectively. In this section, we 
investigate the dependence of the CMF on these parameters. 
Indeed, both are likely varying as one approaches
 a supernova remnant or  a massive star for example.

\subsubsection{Dependence on cosmic rays}
\begin{figure} 
\begin{picture} (0,13)
\put(0,7){\includegraphics[width=9cm]{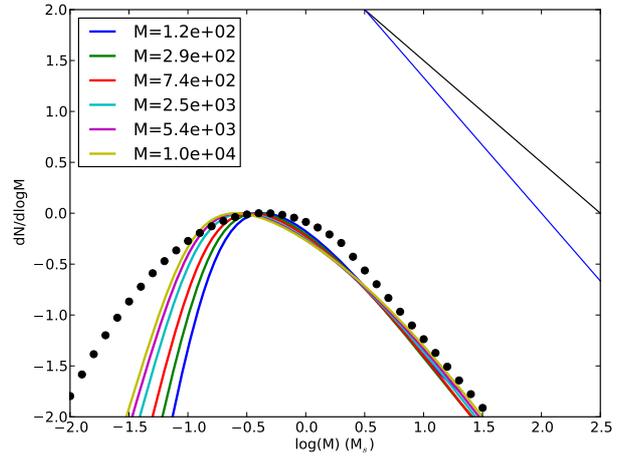}}
\put(0,0){\includegraphics[width=9cm]{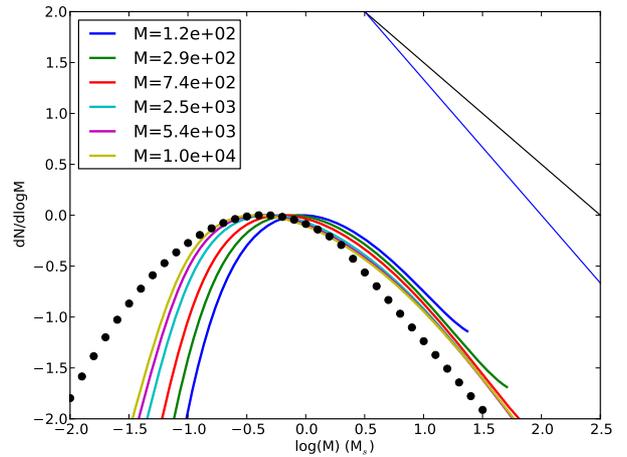}}
\end{picture}
\caption{Same as Fig.~\ref{dN_dlogM} for the fiducial value $\alpha_{*,c}=2$
but for two values of cosmic rays. The upper panel shows 
cosmic ray  heating twice lower
 than the fiducial value while the lower panel 
shows
 cosmic ray heating three times higher. }
\label{dN_dlogM_cos}
\end{figure}

Figure~\ref{dN_dlogM_cos} shows the mass distribution for 
two values of cosmic rays heating, namely half (upper panel)
and three times the fiducial value stated by 
Eq.~(\ref{cosmic_heat}). As can be seen from a comparison 
with the upper panel of Fig.~\ref{dN_dlogM}, 
cosmic ray heating has an impact on the mass distribution that
is only moderate. For most of the distribution, the variations 
are typically less than a factor two.

\subsubsection{Dependence on radiation}
\begin{figure} 
\begin{picture} (0,13)
\put(0,7){\includegraphics[width=9cm]{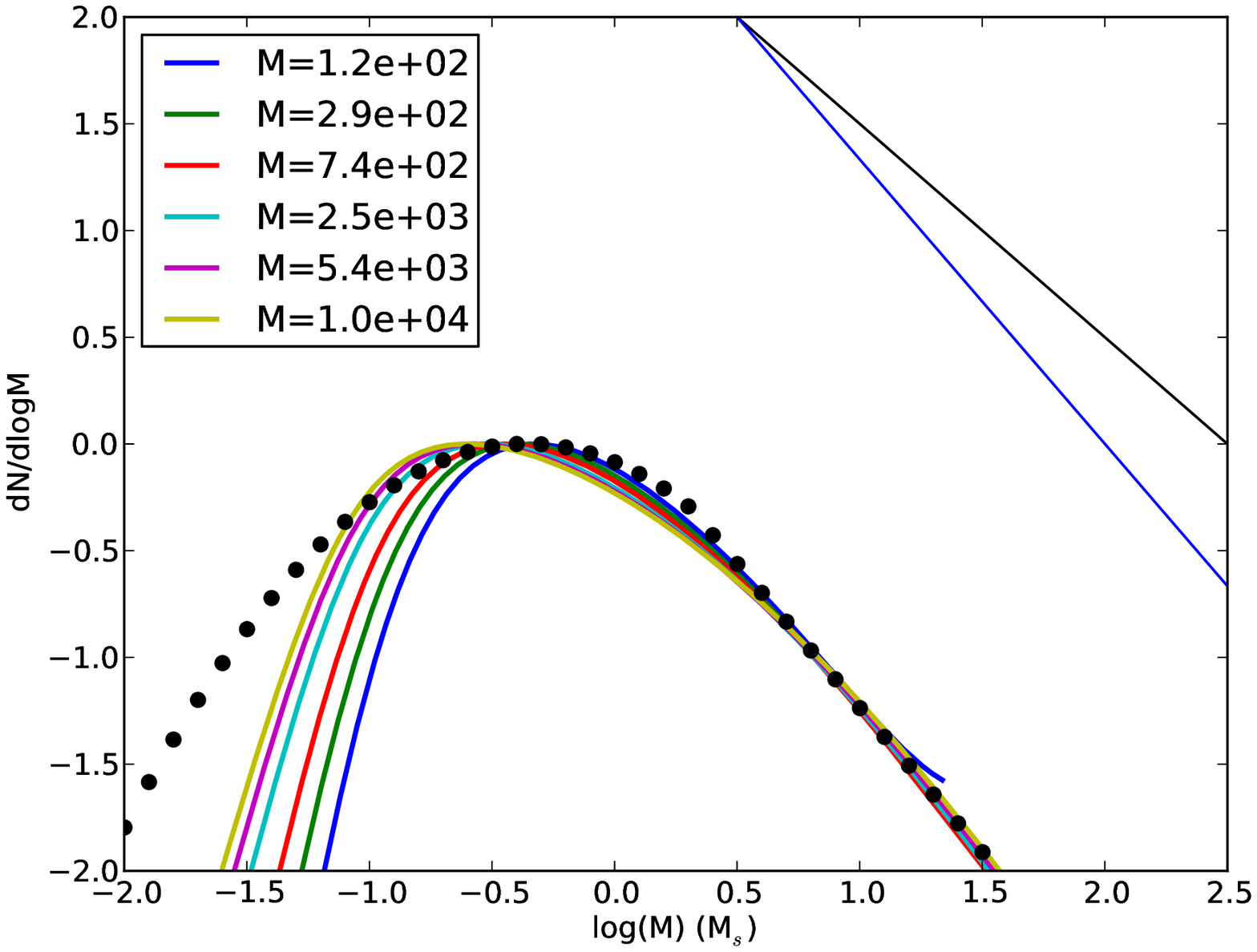}}
\put(0,0){\includegraphics[width=9cm]{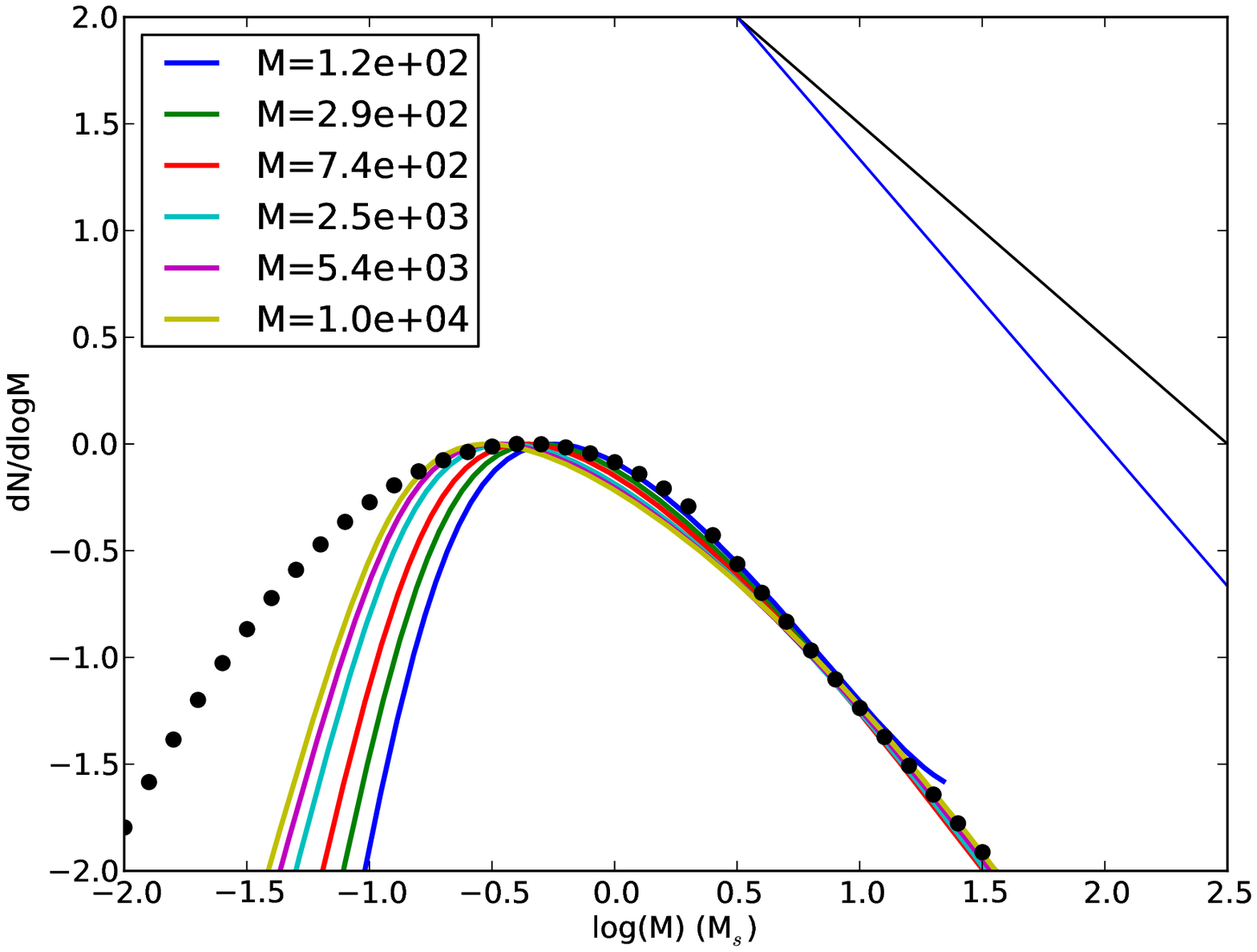}}
\end{picture}
\caption{Same as Fig.~\ref{dN_dlogM} for the fiducial value $\alpha_{*,c}=2$
but for two values of the radiation field. The upper panel shows an 
interstellar radiation field equal to half the canonical values while the
lower panel shows a radiation field twice as high. }
\label{dN_dlogM_rad}
\end{figure}

When increasing the interstellar radiative fields, the various   
components must be distinguished since they have a different 
physical origin. For simplicity we multiply all components 
by the same factor except for the cmb component, which is kept constant.  
Figure~\ref{dN_dlogM_rad} shows the mass distribution for 
two values of interstellar radiation field, namely half (upper panel)
and twice the fiducial value. The departure from the 
mass distribution displayed in Fig.~\ref{dN_dlogM} 
is even smaller than for the cosmic rays. This is because the 
radiation field is modifying  the dust temperature, 
which has only a weak impact on the mass distribution because, 
as discussed above, the effective $\gamma$ at low and 
high density are about 0.85-9 and 1, respectively 
(see Fig.~\ref{dens_temp}).

\subsection{Discussion}
One of the central questions regarding the IMF is that of its apparent universality, that is to say, 
its apparent lack of variations in the different 
determinations that have been made so far (e.g. 
Moraux et al. 2007, Bastian et al. 2010). 
The various estimates in open clusters and young 
clusters lead to some variability on the IMF 
parameters, in particular on the characteristic mass,
or  peak position, but as pointed out 
by de Marci et al. (2010) dynamical evolution is a plausible 
explanation to account for the strongest variations. 
There is  a possible variation independent of the cluster
dynamical evolution, however, as can be seen for example 
in Fig.~3 (left panel) of Bastian et al. (2010) and 
in Fig.~2 of de Marci et al. (2010), which shows that at a given 
dynamical age, the value of $m_c$ typically varies over a
  factor of about 2. It is worth stressing that a possible deficiency of brown 
dwarfs is observed in Taurus (e.g. Guieu et al. 2006, 
Luhman et al. 2009, Bastian et al. 2010)
 while a possible 
excess (factor of about 2) may have been observed in $\sigma$ Ori and
Upper Sco (e.g. Caballero et al. 2007). These
variations need to be confirmed, however.  It is clearly difficult,
 at this stage, to give a definite conclusion.

On the other hand, the results obtained in this paper can be broadly summarized as follows:
$i)$ the high mass part of the IMF is robust to all variations
explored in this paper (higher accretion rate could lead to some 
flattening and shift the peak toward smaller masses), 
$ii)$ the low mass part varies with the mass of the 
cluster and the accretion rate by a factor 
of about 2 
in the range we  explored. The variation with the cluster mass 
in the range 500-10$^4$ $M_\odot$  is very limited, typically less than a factor 2 
(see the four more massive cases displayed in Fig.~\ref{dN_dlogM})
and thus account well for the lack of variability since as 
displayed in table~1 of de Marci et al. (2010), most of the clusters 
that we studied have a mass of stars that is comparable to or greater
than, $10^2$ $M_\odot$. 
This in turn corresponds to a mass of gas about
 five times larger,
assuming an efficiency of 20\%. 
The variation of the mass function with
 the accretion rate, i.e. with $\alpha_{*,c}$, 
is more significant but remains compatible with the range of values
that we  inferred from the comparison with the data from
Lada \& Lada (2003), that is to say a factor of about 2 for the accretion 
rate (at fixed mass) and a factor of about 2-3 on the low mass 
part of the distribution (with much stronger variations for the very low mass 
objects).

\section{Conclusion}

We have developed an analytical model that 
successfully reproduces the mass-size relation 
of embedded clusters. It is based on the 
continuous accretion of gas that drives
the turbulence which in turn resists 
the gravitational contraction and determines
the cluster radius. Comparing the mass-size relation
with the available data of embedded clusters, we showed 
that they can be well fitted by employing an accretion 
rate that is entirely reasonable and close to the rate
that can be inferred using Larson relations. Moreover,
the variation of  this parameter needed to reproduce 
the dispersion of observational data is moderate 
being equal to a factor of about 2. 
Eventually, the turbulent 
energy dissipates and heats the gas. Performing a thermal 
balance, we  calculated the temperature 
distribution within the cluster and  applied 
a time-dependent version of the
 HC2008 theory to obtain the mass spectrum
of self-gravitating condensations. 
The peak position and more generally the whole
mass spectrum is inferred from a large domain of 
proto-cluster masses. We found that for 
gas masses between 500 to $10^4$ $M_\odot$ corresponding
to a mass of stars roughly 5 times smaller, 
the mass spectrum of self-gravitating condensations
does not vary significantly. The mass spectrum 
varies more significantly with the accretion rate,
but given the range of accretion rate deduced
from the observational comparison,  they remain 
compatible with the available determination of the 
IMF in the Galaxy.

\section{Acknowledgments}
We thank the anonymous referee for comments
that have significantly improved the original manuscript.
PH warmly thanks Gilles Chabrier, J\'er\^ome Bouvier 
and Ralf Klessen
for many related stimulating discussions. He also thanks
Eugene Chiang and Chris Matzner for stimulating 
discussions on cluster formation during the ISIMA summer school
hosted by the KIAA in Beijing.

\appendix

\section{Virial theorem for an accreting system}
The virial theorem is derived in many textbooks 
(e.g. Shu 1992, Kulsrud 2005, Lequeux 2005) and does not
need to be presented. However, it is most of the time 
assumed that the system has a constant mass, i.e. is not 
accreting. Here we derive its expression for a system 
 accreting at a rate $\dot{M}$  (see also Goldbaum et al. 
2011). That is to say, at 
radius $R(t)$, gas at density $\rho_{inf}$ is falling 
into the cluster with a velocity $v_{inf}$. 

We start as usual by multiplying the momentum conservation equation 
by the vector position ${\bf r}$ and integrating over the volume. 
The right-hand side of the momentum conservation equation is equal 
to the sum of the forces, namely thermal
pressure and gravity (and Lorentz force when the 
magnetic field is considered). 
It can be rewritten in the unmagnetized case as $2 E_{therm} - 3 P_{therm} V
+E_g$, where  $E_{therm}$ is the thermal energy, $P_{therm}$ is the thermal 
pressure, $V$ is the volume and $E_g$ the gravitational energy.
The left-hand side is given by $\int \rho d_t v_i r_i dV$. 
Integrating this expression by part, we have
\begin{eqnarray}
\nonumber
\int \rho d_t v_i r_i dV &=& \int \rho d_t(v_i r_i) dV - \int \rho v_i^2 dV, \\
&=& \int \rho \partial_t(v_i r_i) dV + \int \rho {\bf v.grad}(v_i r_i) dV \\
\nonumber
& &- \int \rho v_i^2 dV, \\
&=& \partial_t (\int \rho v_i r_i dV) + \int {\rm div} (\rho v_i r_i {\bf v}) dV \nonumber \\ & &- 
\int \rho v_i^2 dV, \nonumber \\
&=& \partial_t (\int \rho d_t(r_i^2) / 2 dV) + \int  \rho v_i r_i {\bf v} {\bf dS} \nonumber \\ & &- 
\int \rho v_i^2 dV, \nonumber
\label{eq1}
\end{eqnarray}
where we  used the continuity equation $\partial_t \rho + {\rm div}(\rho {\bf v})=0$
to cancel some of the terms. 

Let us consider now the first term which appear in the right-hand side. We can write
\begin{eqnarray}
\nonumber
\int \rho v_i r_i  dV &=& \int \rho d_t(r_i^2) / 2 dV = \int \rho \partial_t(r_i^2) / 2 dV  \\ \nonumber & &+
\int \rho {\bf v.grad} (r_i^2 /2 ) dV,  \\
&=& \int \partial_t(\rho r_i^2 / 2) dV + 
\int {\rm div}(\rho r_i^2 {\bf v}) / 2  dV, \\
&=& \partial_t( \int \rho r_i^2 / 2 dV) + \int \rho r_i^2 / 2  {\bf v.dS}.
\nonumber
\label{eq2}
\end{eqnarray}

Thus Eqs.~(\ref{eq1}) and (\ref{eq2}) lead to
\begin{eqnarray}
\nonumber
\int \rho d_t v_i r_i dV &=& 
 \partial_t^2 ( \int \rho r_i^2 / 2 dV ) + \partial_t (\int \rho r_i^2 / 2  {\bf v.dS}) \\ \nonumber & &
 + \int  \rho v_i r_i {\bf v} {\bf dS} - \int \rho v_i^2 dV, \\
&=& {1/2} \partial^2_{t^2} I - \partial_t ( \rho v_{inf} R^2 S) 
\\ \nonumber 
& & +   \rho_{inf} v_{inf}^2 R S - 2 E_c, \\
&=& {1/2} \partial^2_{t^2} I - {1 \over 2} \partial_ t(\dot{M}) R^2  - 
{1 \over 2} \dot{M}^2 \partial_M R^2  \nonumber \\ 
& &+   3 P_{ram} V - 2 E_c, \nonumber
\label{eq3}
\end{eqnarray}
where $I = \int \rho r_i^2 dV$, $\dot{M}=\rho_{inf} v_{inf} S$ and $P_{ram} = \rho_{inf} v_{inf}^2$.
Note that the  minus sign in the second and third terms of the right-hand side, are due 
to the fact that as the gas is infalling, $v_{inf}<0$ and thus ${\bf v.dS}<0$.

The complete expression for the virial theorem is thus 
\begin{eqnarray}
\nonumber
{1/2} \partial^2_{t^2} I &=&  {1 \over 2} \partial_ t(\dot{M}) R^2  + 
{1 \over 2} \dot{M}^2 \partial_M R^2  -   
3 (P_{ram} + P_{therm}) V  \\ 
& & + 2 E_c  +  2 E_{therm} + E_g. 
\end{eqnarray}

Assuming stationarity and neglecting  thermal support and thermal pressure, we derive
\begin{eqnarray}
{1 \over 2} \dot{M}^2 \partial_M R^2  - 3 P_{ram} V  + 2 E_c  + E_g = 0, 
\label{eq_pap}
\end{eqnarray}
which is the equation used in this paper.

\section{Gravitational energy released by accretion}
We calculate here the gravitational energy that is released when 
a particle of gas is accreted onto the proto-cluster. For that purpose,
we assume the following. First, the accreted particles 
that enter the proto-cluster, will on average be distributed 
uniformly within the cluster, second, we assume that the gravitational 
energy inside the cluster is produced by a uniform distribution of matter,
i.e. we neglect the influence of any local density fluctuations.

Under these assumptions
the gravitational force inside the proto-cluster is $-G 4 \pi / 3 \rho_* r$
and the gravitational potential is thus $G 2 \pi /3 ( r^2 - 3 R_*^2 ) $.
Since the gravitational potential outside the cluster is 
$-G M_* / r$, it can easily be checked that the gravitational potential is 
continuous throughout the cluster boundary, i.e. at $r=R_*$.
Therefore we get that the mean specific gravitational energy, $e_g$,
of a particle that falls into the proto-cluster is
\begin{eqnarray}
e_g =  - 
\int _0 ^{R_*} {G 2 \pi \rho_* \over 3} (3 R_*^2 - r^2) f(r) dr = 
-{6 \over 5} {G M_* \over R_*},
\end{eqnarray}
where 
$f(r)dr=3 r^2 dr /R_*^3$ is the fraction of particles located between $r$ and $r+dr$.

Thus the total accretion energy rate is 
\begin{eqnarray}
\dot{E}_{\rm ext}={6 \over 5} {G M_* \dot{M} \over R_*}.
\end{eqnarray}

\end{document}